\documentstyle[preprint,epsf,aps]{revtex}

\tighten
\draft

\begin{document}

\def\ny{{\cal N}_Y} 
\def\f{2\Delta_f}
\def\B{2\Delta_B}
\def\ca{\cosh {{\alpha}\over{2}}}
\def\cb{\cosh^2 {{\alpha}\over{2}}}

\newcommand{\fig}[2]{\epsfxsize=#1\epsfbox{#2}}
\def\Im{\mathop{\hbox{Im}}}
\def\Re{\mathop{\hbox{Re}}}
\parindent=0pt

\title{Overscreened multi-channel $\mbox{SU}(N)$ Kondo model: 
large-N solution and conformal field theory}
\author{Olivier Parcollet, Antoine Georges} 
\address{ Laboratoire de Physique Th{\'e}orique de l'Ecole 
Normale Sup{\'e}rieure,\cite{auth1}
24 rue Lhomond, 75231 Paris Cedex 05, France}
\author{Gabriel Kotliar, Anirvan Sengupta}
\address
{Serin Physics Laboratory, Rutgers University,
Piscataway, NJ 08854, USA}
\date{\today}
\maketitle
\begin{abstract}
The multichannel Kondo model with 
$\mbox{SU}(N)$ spin symmetry and $\mbox{SU}(K)$ 
channel symmetry is considered. The impurity spin is chosen to 
transform as an 
antisymmetric representation of $\mbox{SU}(N)$, corresponding to a 
fixed number of Abrikosov fermions $\sum_{\alpha}
f_{\alpha}^{\dagger}f_{\alpha}=Q$. For more than one channel 
($K>1$), and all values of $N$ and $Q$, the model displays non-Fermi 
behaviour associated with the {\it overscreening} of the impurity spin. 
Universal low-temperature thermodynamic 
and transport properties of this non-Fermi liquid state 
are computed using conformal field theory methods. 
A large-$N$ limit of the model is then considered, in which 
$K/N\equiv\gamma$ and $Q/N\equiv q_0$ are held fixed. 
Spectral densities satisfy coupled integral equations in this  
limit, corresponding to a (time-dependent) 
saddle-point. A low frequency, low-temperature analysis of these equations 
reveals universal scaling properties in the variable $\omega/T$, in agreement 
with conformal invariance. The universal scaling form is obtained 
analytically and used to compute  the low-temperature 
universal properties of the model in the large-N limit, such as 
the $T=0$ residual entropy and residual resistivity, and the 
critical exponents associated with the specific heat and susceptibility. 
The connections with the ``non-crossing approximation'' and the 
previous work of Cox and Ruckenstein are discussed.  
 
\end{abstract}

\pacs{PACS numbers: 75.20 Hr, 75.30 Mb, 71.10 Hf. Preprint LPTENS : 97/55 }


\section{Introduction and Model}

Multichannel Kondo impurity models \cite{NB,CZ} 
have recently attracted considerable 
attention, for several reasons. First, in the overscreened case, 
they provide an explicit example of a non-Fermi liquid ground-state. 
Second, these models can be studied by a variety of controlled techniques, 
and provide invaluable testing grounds for theoretical methods 
dealing with correlated electron systems. One of the most recent 
and fruitful development 
in this respect has been the conformal field-theory 
approach developed by Affleck and Ludwig \cite{AL,AL2,ALent}. Finally, 
multichannel models have experimental relevance to tunneling phenomena 
in quantum dots and two level systems \cite{Moriond}, and possibly 
also to some heavy-fermion compounds \cite{CZ}. 

In this paper, we consider  
a generalisation of the multi-channel Kondo model, in which 
the spin symmetry group is extended from $\mbox{SU}(2)$ to 
$\mbox{SU}(N)$. In addition, the model has a $\mbox{SU}(K)$ symmetry 
among the $K$ ``channels'' (flavours) of conduction electrons. 
We focus here on spin representations which are such that the 
model is in the non-Fermi liquid {\it overscreened} regime. 
We shall derive in this paper several universal properties of this 
$\mbox{SU}(N)\otimes\mbox{SU}(K)$ Kondo model  
in the low-temperature regime. Specifically, we obtain: 
the zero-temperature residual entropy, the zero-temperature 
impurity resistivity and T-matrix, and the critical exponents 
governing the leading low-temperature behaviour of the impurity 
specific heat, susceptibility and resistivity.   
   
These results will be obtained using two different approaches. It is 
one of the main motivations of this paper to compare these two 
approaches in some detail. 
First (Sec.\ref{CFTsec}), we apply CFT methods to study the model 
for general values of  $N$ and $K$. Then, we study the limit of 
large $N$ and $K$, with $K/N=\gamma$ fixed. This limit was previously 
considered by Cox and Ruckenstein \cite{CR}, in connection with the 
``non-crossing approximation'' (NCA) \label{NCA}. 
There is a crucial difference 
between our work and that of Ref.\cite{CR} however, which is that 
we keep track of the quantum number specifying the spin representation of 
the impurity by imposing a constraint (on the Abrikosov fermions 
representing the impurity) which also scales proportionally to $N$. 
As a result, the solution of the model at large $N$ follows from a 
true saddle-point principle, with controllable fluctuations in $1/N$.
Hence, a detailed quantitative comparison of 
the large-N limit to the CFT results can be made. 
The saddle-point equations are coupled integral equations similar in 
structure to those of the NCA, except for the different handling of the 
constraint. The $T=0$ impurity 
entropy and residual resistivity are obtained for the first time 
in analytical form in this paper from a low-energy analysis of these 
coupled integral equations and shown to agree with the large-N 
limit of the CFT results. We also demonstrate that the spectral 
functions resulting from these equations take a universal scaling 
form in the limit 
$\omega, T \rightarrow 0$, which is precisely that expected from 
the conformal invariance of the problem.

The Hamiltonian of the model considered in this paper reads:
\begin{equation}
\label{ham1}
H\, =\, \sum_{\vec{p}}\sum_{i=1}^{K}\sum_{\alpha=1}^{N} 
\epsilon(\vec{p}) c^\dagger_{\vec{p}i\alpha} c_{\vec{p}i\alpha}
\,+\,J_{K} \sum_{A=1}^{N^2-1} S^A 
\sum_{\vec{p}\vec{p'}i\alpha\beta} 
c^{\dagger}_{\vec{p}i\alpha} t^A_{\alpha\beta} c_{\vec{p'}i\beta}
\end{equation} 
In this expression, $c^\dagger_{\vec{p}i\alpha}$ creates an electron in the 
conduction band, with momentum $\vec{p}$, channel (flavour) index 
$i=1,\cdots,K$ and $\mbox{SU}(N)$ spin index $\alpha=1,\cdots,N$. 
The conduction electrons transform under the fundamental representation of 
the $\mbox{SU}(N)$ group, with generators $t^A_{\alpha\beta}$ 
($A=1,\cdots,N^2-1$). They interact with a localised spin degree of freedom 
placed at the origin, $\vec{S}=\{S^A, A=1,\cdots,N^2-1\}$ which is assumed 
to transform under a given irreducible representation $R$ of the 
$\mbox{SU}(N)$ group. 

In the one-channel case ($K=1$), and when $R$ is taken to be the 
fundamental representation, this is the Coqblin-Schrieffer model 
of a conduction gas interacting with a localised atomic level with 
angular momentum $j$ ($N=2j+1$) \cite{CS}. 
In this article, we are interested in the 
possible non-Fermi liquid behaviour associated with the multi-channel 
generalisation ($K>1$) \cite{NB} \cite{CZ}.
We shall mostly focus on the case where $R$ corresponds to 
antisymmetric tensors of $Q$ indices, {\it i.e} the Young tableau 
associated with $R$ is made of a {\sl single column} of $Q$ indices. 
In that case, it is convenient to use an explicit representation of the 
localised spin in terms of $N$ species of auxiliary fermions 
$f_{\alpha}$ ($\alpha=1,\cdots,N$), constrained to obey: 
\begin{equation}
\label{constr}
\sum_{\alpha=1}^{N} f^\dagger_{\alpha}f_{\alpha} \,=\, Q
\end{equation}
so that the $N^2-1$ (traceless) 
components of $\vec{S}$ can be represented as: 
$S_{\alpha\beta} = f^+_{\alpha}f_{\beta}- {Q\over N} \delta_{\alpha\beta}$.
For these choices of $R$, the 
Hamiltonian can be written as (after a reshuffling of indices using a Fierz
identity ):
\begin{equation}
\label{ham2}
H\, =\, \sum_{\vec{p}}\sum_{i=1}^{K}\sum_{\alpha=1}^{N}
\epsilon(\vec{p}) c^\dagger_{\vec{p}i\alpha} c_{\vec{p}i\alpha}
\,+\,J_{K} \sum_{\vec{p}\vec{p'}i\alpha\beta} 
(f^\dagger_{\alpha}f_{\beta}-{Q\over N} \delta_{\alpha\beta}) 
\, { c^\dagger_{\vec{p}i\beta} c_{\vec{p'}i\alpha}}
\end{equation}
In a recent paper \cite{OPAG}, the case of a {\it symmetric} representation 
of the impurity spin (corresponding to a Young tableau made of a single 
{\it line} of $P$ boxes) has been considered by two of us. In that case, 
a transition from overscreening to underscreening is found as a function of 
the ``size'' $P$ of the impurity spin. In contrast, the antisymmetric 
representations considered in the present paper always lead to overscreening 
(except for $K=1$ which is exactly screened), as shown below. As long as 
only the  overscreened regime is considered, 
the analysis of the present paper applies to 
symmetric representations as well, up to some straightforward 
replacements.     

\section{Strong coupling analysis}
\label{strongcoupling}
It is easily checked that a weak antiferromagnetic coupling ($J_K>0$) 
grows under renormalisation for all $K$ and $N$, and all representations  
$R$ of the local spin. What is needed is a physical argument in order to 
determine whether the R.G flow takes $J_K$ all the way to strong coupling 
(underscreened or exactly screened cases), or whether an intermediate 
non-Fermi liquid fixed point exists (overscreened cases). 

Following the Nozieres and Blandin \cite{NB} analysis of the multichannel 
$\mbox{SU}(2)$ model, we consider the strong-coupling fixed point 
$J_K=+\infty$. In this limit, the impurity spin binds a certain number 
of conduction electrons (at most $NK$ because of the Pauli principle). 
The resulting bound-state corresponds to  
a new spin representation $R_{sc}$ which is dictated by the 
minimisation of the Kondo energy. For the specific choices of $R$ above, 
we have proven that:

\begin{itemize}
\item In the one-channel case ($K=1$) and for arbitrary $N$ and $Q$, 
$R_{sc}$ is the ``singlet'' representation (of dimension 
$d(R_{sc})=1$). It is obtained by binding $N-Q$ conduction electrons to the 
$Q$ pseudo-fermions $f_{\alpha}$. The 
impurity spin is thus {\sl exactly screened} at strong-coupling.

\item For all multi-channel cases ($K\geq 2$, arbitrary $N$ and $Q$), the 
ground-state  at the strong-coupling fixed point is the representation 
$R_{sc}$ characterised by a rectangular Young tableau with $N-Q$ lines 
and $K-1$ columns. Its dimension $d(R_{sc})$ ({\it i.e} the degeneracy of 
the strong-coupling bound-state) is larger than the degeneracy at zero 
coupling, given by $d(R)=$ $N \choose Q$ $\equiv N!/Q!(N-Q)!$.

\end{itemize}

The Young-tableau associated with the strong-coupling state in both cases 
is depicted in Fig.\ref{drawsc}. 
The detailed proof of these statements and the explicit construction of 
$R_{sc}$ are given in Appendix \ref{appendixstrongcoupling}.
These properties are sufficient to establish that :
\begin{itemize}
\item In the one-channel case, the strong coupling fixed point is stable 
under R.G., and hence the impurity spin is exactly screened by the Kondo 
effect.
\item  In the multi-channel case, a direct R.G flow from 
weak to strong-coupling is impossible, thereby suggesting the existence 
of an intermediate coupling fixed-point (``overscreening''). 
\end{itemize}

The connection between these statements and the above results on the 
nature and degeneracy of the strong-coupling bound-state is clear on
 physical grounds. Indeed, it is not possible to flow under renormalisation 
from a fixed point with a lower ground-state degeneracy to a 
fixed-point with a higher one, because the effective number of degrees 
of freedom can only decrease under R.G. Hence, no flow away from the 
strong-coupling fixed point is possible in the one-channel ($K=1$)
case since 
the strong-coupling state is nondegenerate. Also, no direct flow from 
weak to strong coupling is possible for $K\geq  2$ since $d(R_{sc})>d(R)$. 
These statements can be made more rigorous \cite{ALent} by considering 
the impurity entropy defined as:
\begin{equation}
\label{defentropy}
S_{imp} \equiv \lim_{T\rightarrow 0} \lim_{V\rightarrow\infty} 
[S(T)-S_{bulk}(T)]
\end{equation}
Where $S_{bulk}$ denotes the contribution to the entropy which is 
proportional to the volume $V$ (and is simply the contribution of the 
conduction electron gas), and care has been taken in specifying the order 
of the infinite-volume and zero-temperature limits. 
At the weak coupling fixed point $S_{imp}(J_K=0) = \mbox{ln}\,d(R)$, while 
$S_{imp}(J_K=\infty) = \mbox{ln}\,d(R_{sc})$ at strong-coupling.  
$S_{imp}$ must decrease under renormalisation \cite{ALent}, a property 
which is the analogue for boundary critical phenomena to Zamolodchikov's 
``c-theorem'' in the bulk. This suggests a R.G flow of the kind 
indicated above. This conclusion can of course be confirmed by a 
perturbative calculation (in the hopping amplitude) around the 
strong-coupling fixed-point \cite{NB}.

The value of $S_{imp}$ will be calculated below  at the intermediate 
fixed-point in the overscreened case, and found to be non-integer (as in the 
$N=2$ case \cite{ALent}).
 
\section{Conformal field theory approach}
\label{CFTsec}
Having established the existence of an intermediate fixed-point for 
$K \geq 2$, we sketch some of its properties that can be obtained from 
conformal field-theory (CFT) methods. This is a straightforward 
extension to the $\mbox{SU}(N)$ case of Affleck and Ludwig's approach for 
$\mbox{SU}(2)$ \cite{AL,AL2}. The aim of this section is not to present
 a complete 
conformal field-theory solution, but simply to derive those 
properties which will be compared with the large-N explicit solution 
given below.
 
In the CFT approach, the model (\ref{ham1}) is first mapped at low-energy 
onto a $1+1$-dimensional model of $NK$ chiral fermions. At a fixed point, 
this model has a local conformal symmetry based on the Kac-Moody algebra 
$\widehat{SU}_K(N)_{s}\otimes \widehat{SU}_N(K)_{f}\otimes
\widehat{U}(1)_{c}$ corresponding to the spin-flavour-charge decomposition 
of the degrees of freedom. The free-fermion spectrum at the weak-coupling 
fixed point can be organised in multiplets of this symmetry algebra: to each 
level corresponds a primary operator in the spin, flavour and charge
 sectors. 
A major insight \cite{AL} is then that the spectrum at the infra-red stable, 
intermediate coupling fixed point can be obtained from a
 ``fusion principle''.
Specifically, the spectrum is obtained by acting, in the spin sector, 
on the primary operator associated with a given free-fermion state,  
with the primary operator of the $\widehat{SU}_K(N)_{s}$ algebra 
corresponding to the representation $R$ of the 
impurity spin (leaving unchanged the flavour and charge sectors). 
The ``fusion rules'' of the algebra determine the new operators associated 
with each energy level at the intermediate fixed point. 

This fusion principle 
also relates the impurity entropy $S_{imp}$ as defined above to the 
``modular S-matrix'' $S_{0}^{R}$ of the $\widehat{SU}_K(N)$ algebra 
(this is the matrix which specifies the action of a modular transformation 
on the irreducible characters of the algebra corresponding to a given 
irreducible representation $R$). Specifically, denoting by $R=0$ the 
trivial (identity) representation: 
\begin{equation}
S_{imp}\,=\, \mbox{ln} {{S_{0}^{R}}\over{S_{0}^{0}}}
\end{equation}
The expression of the modular S-matrix for $\widehat{SU}_K(N)$ can be 
found in the litterature \cite{Smodulaire}.
We have found particularly useful to make use of an elegant formulation 
introduced by Douglas \cite{Douglas}, which is briefly explained in 
Appendices \ref{appendixstrongcoupling} and \ref{modsmatrix}.  
For the representations $R$ associated to a single column of 
length $Q$ (corresponding to (\ref{constr})), one finds using this
 representation:
\begin{equation}
\label{entropieImpureteCFT}
S_{imp}\,=\, \mbox{ln}\,\prod_{n=1}^{Q} 
{{\sin {{\pi (N+1-n)}\over{N+K}}}\over{\sin {{\pi n}\over{N+K}}}}
\end{equation}
It is easily checked that indeed $S_{imp} < \mbox{ln} d(R)$ for all values 
of $N$, $Q$ and $K$. Note also that this expression correctly yields 
$S_{imp}=0$ in the exactly screened case $K=1$ (for arbitrary $N,Q$).

The low-temperature behaviour of various physical quantities can also be 
obtained from the CFT approach. At the intermediate coupling fixed 
point, the local impurity spin acquires the scaling dimension of the 
primary operator of the $\widehat{SU}_K(N)_s$ algebra associated with 
the ($N^2-1$ dimensional) adjoint representation of $\mbox{SU}(N)$. 
Its conformal dimension $\Delta_s$ (such that 
$<S(0)S(t)>\sim 1/t^{2\Delta_s}$ at $T=0$) reads: 
\begin{equation}
\label{spindim}
\Delta_s\,=\, {{N}\over{N+K}}
\end{equation}
Integrating this correlation function, this implies that the 
{\it local susceptibility} 
$\chi_{loc}\propto \int_{\tau_0}^{1/T} <S(0)S(\tau)> d\tau$ 
(corresponding to the coupling of  
an external field to the impurity spin {\it only}) diverges at low 
temperature when $K\geq N$, while it remains finite for 
$K<N$:
\begin{eqnarray}
\label{chiloc}
\nonumber & K\geq N\,\,\,:\,\,\, \chi_{loc}\sim \left({{1}\over{T}}\right)
^{(K-N)/(K+N)}\\
\nonumber & K=N\,\,\,:\,\,\, \chi_{loc}\sim \mbox{ln} 1/T\\
& K<N\,\,\,:\,\,\, \chi_{loc}\sim \mbox{const.}
\end{eqnarray}
Exactly at the fixed point, the singular contributions to the 
specific heat and {\it impurity susceptibility} 
$\chi_{imp}=\chi-\chi_{bulk}$ (defined by coupling a magnetic field to the 
total spin density) vanish \cite{AL,EKetc}. 
Indeed, the singular behaviour is controlled  
by the leading irrelevant operator compatible with all 
symmetries that can be generated. An obvious candidate for this operator is 
the spin, flavour and charge singlet obtained by contracting the spin 
current with the adjoint primary operator above, 
$\vec{J}_{-1}\cdot\vec{\phi}$. It has dimension 
$1+{{N}\over{N+K}} = 1+\Delta_s$. 
This leads to a singular contribution 
to the impurity susceptibility (arising from perturbation 
theory at {\it second order} in the irrelevant operator \cite{AL,EKetc}) 
of the same nature than for $\chi_{loc}$ 
given in (\ref{chiloc}): $\chi_{imp}\sim\chi_{loc}$. 

Another irrelevant operator 
can be constructed in the flavour sector in an analogous manner, namely 
$\vec{J}_{-1}^f\cdot\vec{\phi^f}$. This operator has dimension 
$K/(N+K)$, which is thus lower than the above operator in the spin sector 
when $K<N$. This operator could a priori contribute to the 
low-temperature behaviour of the specific heat, which would lead to a 
divergent specific heat coefficient $C/T$ for both $K>N$ and $K<N$. 
On the basis of the explicit large-N calculation presented below, we believe 
however that this operator is {\it not generated} for model (\ref{ham1}) when 
the conduction band d.o.s is taken to be perfectly flat and the 
cutoff is taken to infinity (conformal limit), so that the specific 
heat ratio has the same behaviour than the susceptibilities above: 
$C/T\sim \chi_{loc}\sim\chi_{imp}$. 
If the model is extended to an impurity spin with internal flavour 
degrees of freedom, this operator will however show up (leading to 
$C/T \sim T^{-(N-K)/(N+K)}$ for $K<N$). 
It also appears (see below) if an Anderson 
model generalisation of model (\ref{ham1}) is considered away from 
particle-hole symmetry. 
  
\section{Saddle-point equations in the large-N limit}

We now turn to the analysis of the large-N limit of this model. 
This will be done by setting:
\begin{equation}
K\,=\,N\gamma \,\,\,,\,\,\,J_K\,= {{J}\over{N}}
\end{equation}
and taking the limit $N\rightarrow\infty$ for fixed values of 
$\gamma$, and $J$, so that the number of channels is also taken 
to be large. 
In Ref.\cite{CR} (see also \cite{CZ}), Cox and Ruckenstein 
considered this limit while holding $Q$ fixed ($Q=1$). 
They obtained in this limit identical results to those of 
the ``non-crossing approximation'' (NCA) \cite{NCA}.
Here, we shall proceed in a different manner by {\it taking $Q$ to be 
large as well}:
\begin{equation}
Q\,=\,q_0 N
\end{equation}
This insures that a true saddle-point exists, with controllable fluctuations 
order by order in $1/N$. It will also allow us to study the dependence 
on the representation $R$ of the local spin, parametrised by $q_0$ 
\cite{foot1}. 
The approach of Ref.\cite{CR} is recovered  in the limit $q_0\rightarrow 0$ 
(or $1$). 

The action corresponding to the functional integral formulation of 
model (\ref{ham2}) reads:
\begin{eqnarray}
\label{action1}
\nonumber &S\,=\,
-\int^{\beta}_{0} d\tau \int^{\beta}_{0} d\tau' \sum_{i\alpha}
c^{\dagger}_{i\alpha}(\tau) G_{0}^{-1}(\tau-\tau') c_{i\alpha}(\tau')\\
\nonumber &+\int^{\beta}_{0} d\tau \sum_{\alpha} 
f^{\dagger}_{\alpha}(\tau) \partial_{\tau} f_{\alpha}(\tau)
+\int^{\beta}_{0} d\tau\, i\mu(\tau) (\sum_{\alpha}
f^{\dagger}_{\alpha}(\tau)f_{\alpha}(\tau)
-q_0 N)\\ 
&+ {{J}\over{N}} \int^{\beta}_{0} d\tau  \sum_{i\alpha\beta} 
c^{\dagger}_{i\alpha}(\tau) c_{i\beta}(\tau)
\left(f^{\dagger}_{\beta}(\tau)f_{\alpha}(\tau ) - 
q_{0}\delta_{\alpha \beta }  \right)
\end{eqnarray}
In this expression, the conduction electrons have been 
integrated out in the bulk, keeping only degrees of freedom at the impurity 
site. $G_0(i\omega_n)\equiv\sum_{\vec{p}} 1/(i\omega_n-\epsilon_{\vec{p}})$ 
is the on-site Green's function associated with the conduction electron 
bath. 
In order to decouple the Kondo interaction, an auxiliary bosonic field 
$B_i(\tau)$ is introduced in each channel, and the conduction electrons 
can be integrated out, leaving us with the effective action:
\begin{eqnarray}
\label{action2}
\nonumber &S_{eff}\,=\,
\int^{\beta}_{0} d\tau \sum_{\alpha}
f^{\dagger}_{\alpha}(\tau) \partial_{\tau} f_{\alpha}(\tau)
+\int^{\beta}_{0} d\tau\, i\mu(\tau) 
(\sum_{\alpha} f^{\dagger}_{\alpha}(\tau)f_{\alpha}(\tau) - q_0 N)
+ {{1}\over{J}} \int^{\beta}_{0} d\tau \sum_i B_i^\dagger B_i \\
& + {{1}\over{N}} \int^{\beta}_{0} d\tau \int^{\beta}_{0} d\tau' 
\sum_{i\alpha} 
B_i(\tau) f^{\dagger}_{\alpha}(\tau) G_{0}(\tau-\tau') 
B_i^\dagger(\tau') f_{\alpha}(\tau')
\end{eqnarray}

The quartic term in this expression can be decoupled formally using
two bi-local fields $Q(\tau,\tau')$ and $\overline{Q}(\tau,\tau')$
conjugate to
$\sum_{i} B_i^\dagger(\tau') B_i(\tau)$ and 
$\sum_{\alpha}f^{\dagger}_{\alpha}(\tau)f_{\alpha}(\tau')$ respectively, 
leading to the action :
\begin{eqnarray}
\nonumber S\,&&=\,
\int^{\beta}_{0} d\tau \sum_{\alpha}
f^{\dagger}_{\alpha}(\tau) \partial_{\tau} f_{\alpha}(\tau)
+ {{1}\over{J}} \int^{\beta}_{0} d\tau \sum_i B_i^\dagger B_i 
+\int^{\beta}_{0} d\tau\, i\mu(\tau)
(\sum_{\alpha} f^{\dagger}_{\alpha}(\tau)f_{\alpha}(\tau) - q_0 N)\\
\nonumber &&
-N \int\!\!\! \int d\tau d\tau'\, 
 \overline{Q}(\tau,\tau') G_0^{-1}(\tau-\tau') Q(\tau,\tau') -
 \int\!\!\! \int d\tau d\tau'\, Q(\tau',\tau) \sum_i B_i^+(\tau)B_i(\tau') \\
&& - \int\!\!\! \int d\tau d\tau'\,\overline{Q}(\tau,\tau')
 \sum_{\alpha} f_{\alpha}^+(\tau)f_{\alpha}(\tau')
\end{eqnarray}
The $B$ and $f$ fields can now be integrated out to yield:
\begin{eqnarray}
\nonumber S\,&&=\,
-N \int\!\!\!\int  d\tau d\tau'\, 
 \overline{Q}(\tau,\tau') G_0^{-1}(\tau-\tau') Q(\tau,\tau')
-N q_0 \int d\tau i\mu(\tau) \\
&&
-N \mbox{Tr}  \ln [ (-\partial_{\tau} - i\mu(\tau))\delta(\tau-\tau') + 
\overline{Q}(\tau,\tau')] 
+K \mbox{Tr} \ln [ {{1}\over{J}} \delta(\tau-\tau') - Q(\tau',\tau) ]
\end{eqnarray}
This final form of the action involves only the three fields 
$Q$, $\overline{Q}$ and $\mu$, and scales globally as $N$ thanks to the 
scalings $K=\gamma N$ and $Q=q_0 N$. 
Hence, it can be solved by the saddle-point method in the large-N limit. 
At the saddle-point, 
$\mu_{sp}(\tau)=i\lambda$ becomes static and purely imaginary, 
while $Q_{sp}=Q(\tau-\tau')$ and 
$\overline{Q}_{sp}=\overline{Q}(\tau-\tau')$ {\it retain time-dependence} 
but depend only on the time difference $\tau-\tau'$ (they identify with the 
bosonic and fermionic self-energies, respectively). 

The final form of the coupled saddle-point equations for the fermionic and 
bosonic Green's functions 
$G_f(\tau)\equiv -<Tf(\tau)f^\dagger(0)>$, $G_B(\tau)\equiv 
<TB(\tau)B^\dagger(0)>$ 
and for the Lagrange multiplier field read:
\begin{equation}
\label{sp}
\Sigma_f(\tau)=\gamma G_0(\tau) G_B(\tau)\,\,\,,\,\,\,
\Sigma_B(\tau)= G_0(\tau) G_f(\tau)
\end{equation}
where the self-energies $\Sigma_f$ and $\Sigma_B$ are defined by:
\begin{equation}
\label{defsigma}
G_f^{-1}(i\omega_n) = i\omega_n+\lambda-\Sigma_f(i\omega_n)
\,\,\,,\,\,\,
G_B^{-1}(i\nu_n) = {{1}\over{J}}-\Sigma_B(i\nu_n)
\end{equation}
In these expressions $\omega_n=(2n+1)\pi/\beta$ and $\nu_n=2n\pi/\beta$ 
denote fermionic and bosonic Matsubara frequencies. 
Let us note that the field $B(\tau)$ is simply a commuting auxiliary 
field rather than a true boson (its equal-time commutator vanishes). As a 
result $G_{B}(\tau)$ is a $\beta-$periodic function, but does not share   
the usual other properties of a bosonic Green function (in particular 
at high frequency) . 

Finally, $\lambda$ is determined by the third saddle point
equation :
\begin{equation}
\label{eqlambda}
G_f(\tau=0^-)\equiv {1 \over\beta  }
\sum_n G_f (i\omega_n) e^{i\omega_n 0^+}\,=\,q_0
\end{equation}
These equations are identical in structure to the usual NCA equations, 
except for the last equation (\ref{eqlambda}) 
which implements the constraint 
and allows us to keep track of the choice of representation for the 
impurity spin. 

\section{Scaling analysis at low frequency and temperature}

\subsection{General considerations}
\label{spAnalysis1}
The analysis of the NCA equations in Ref.\cite{MH} 
can be applied in order to find the 
behaviour of the Green's functions in the low temperature, long time 
regime defined by: $T_{K}^{-1}\ll\tau\ll\beta\rightarrow\infty$ (where 
$T_{K}$ is the Kondo temperature). In this regime, a power-law decay of the 
Green's functions is found:
\begin{equation}
\label{longtime}
G_f(\tau)\sim {{A_f}\over{\tau^{2\Delta_f}}}\,\,\,,\,\,\,
G_B(\tau)\sim {{A_B}\over{\tau^{2\Delta_B}}}\,\,\,,\,\,\,
(T_{K}^{-1}\ll\tau\ll\beta\rightarrow\infty)
\end{equation}
The scaling dimensions $2\Delta_f$ and $2\Delta_B$ can be 
determined explicitly by inserting this form into the above saddle-point 
equations and making a low-frequency analysis, as explained in 
Appendix \ref{detailcalcul}. This yields: 
\begin{equation}
\label{exponents}
2\Delta_f = {{1}\over{1+\gamma}}\,\,\,,\,\,\,
2\Delta_B = {{\gamma}\over{1+\gamma}}
\end{equation}
The overall consistency of (\ref{sp},\ref{defsigma}) at large time
also constrains the product of amplitudes $A_fA_B$ 
(Eq.(\ref{prodamplitude}) in Appendix \ref{detailcalcul}) and 
dictates the behaviour of the self-energies (denoting by
$\rho_0=-\mbox{Im}G_0(i0^+)/\pi$ the conduction bath
density of states at the Fermi level) :
\begin{equation}
\Sigma_f(\tau)\sim
\gamma\rho_0 A_B\left({{1}\over{\tau}}\right)^{2\Delta_B+1}\,\,\,,\,\,\,
\Sigma_B(\tau)\sim A_f\rho_0\left({{1}\over{\tau}}\right)^{2\Delta_f+1}
\end{equation}
together with the sum rule:
\begin{equation}
\Sigma_B(\omega=0,\beta=\infty) = {{1}\over{J}}
\end{equation}
The expression (\ref{exponents}) of the scaling dimensions 
$\Delta_f$ and $\Delta_B$ 
is in complete agreement with the CFT result. Indeed, the fermionic field 
transforms as the fundamental representation of the  
$\widehat{SU}(N)_K$ spin algebra, while the auxiliary bosonic 
field transforms as the fundamental representation of the 
$\widehat{SU}(K)_N$ flavour algebra, leading to:
$2\Delta_f={{N^2-1}\over{N(N+K)}}$ and 
$2\Delta_B={{K^2-1}\over{K(N+K)}}$, which agree with 
(\ref{exponents}) in the large-N limit.   
Also, the local impurity spin correlation function, given in the 
large-N limit by   
$\langle S(0)S(\tau)\rangle \propto G_f(\tau)G_f(-\tau)$ is found to 
decay as $1/\tau^{2\Delta_s}$, 
with $\Delta_s=2\Delta_f=1/(1+\gamma)$ in agreement with the CFT result 
(\ref{spindim}).
 
As will be shown below however, these asymptotic behaviours at $T=0$ do 
not provide enough information to allow for  
the computation of the low-temperature behaviour of the impurity 
free-energy and to determine the $T=0$ impurity entropy (\ref{defentropy}). 
One actually needs to determine the Green's functions in 
the limit $\tau,\beta\rightarrow\infty$, but for an arbitrary 
value of the ratio $\tau/\beta$ ({\it i.e} to analyze the low-temperature, 
low-frequency behaviour of (\ref{sp},\ref{defsigma}) keeping the ratio 
$\omega/T$ fixed \cite{Subir}). 
It is easily seen that in this limit the Green's 
functions and their associated spectral densities 
$\rho_{f,B}(\omega)\equiv -{{1}\over{\pi}}\mbox{Im}G_{f,B}(\omega+i0^+)$ 
obey a {\it scaling behaviour} :
\begin{mathletters}
\begin{equation}
\label{scalg}
G_{f,B}(\tau) = A_{f,B} \beta^{-2\Delta_{f,B}}
\,g_{f,B}\left({\tau\over\beta}\right)
 \,\,\,\,\,\,
(T_{K}^{-1}\ll\tau\,,\beta\,\,;\,\,\tau/\beta\,\,\,\mbox{arbitrary})
\end{equation}
\begin{equation}
\label{scalrho}
\rho_f(\omega) = A_f T^{2\Delta_f-1}\, 
\phi_f\left({{\omega}\over{T}}\right)\,\,\,,\,\,\,
\rho_B(\omega) = A_B T^{2\Delta_B-1}\, \phi_B\left({{\omega}\over{T}}\right)
\end{equation}
\end{mathletters}
In these expressions, $g_{f,B}$ and $\phi_{f,B}$ are {\it universal scaling 
functions} which depend only on $\gamma$ and $q_0$ and {\it not on the 
specific shape of the conduction band or the cutoff}.
These scaling functions will now be found in explicit form. 

\subsection{The particle-hole symmetric representation $q_0=1/2$}

We shall first discuss 
the case where the representation $R$ has $Q=N/2$ boxes ($q_0=1/2$), for   
which there is a particle-hole symmetry among pseudo-fermions 
under $f^\dagger_{\alpha}\leftrightarrow f_{\alpha}$. 
The expression of the scaling functions $g_{f,B}$ in that case can be easily 
guessed from general principles of conformal invariance. 
The idea is that, 
in the limit $T_{K}^{-1}\ll\beta$ with $\tau/\beta$ fixed, 
the finite-temperature Green's function can be obtained from the 
$T=0$ Green's function by applying the conformal transformation 
$z=\exp(i2\pi\tau/\beta)$ \cite{Tsvelik}. Applying this to 
the $T=0$ power-law decay given by (\ref{longtime}), one obtains the 
well-known result for the scaling functions 
($\tilde{\tau}\equiv \tau / \beta$):   
\begin{equation}
\label{cftscaling}
g_f(\tilde{\tau};q_0=1/2)= -
\left({{\pi}\over{\sin\pi \tilde{\tau}}}\right)^{2\Delta_f}
\,\,\,,\,\,\,
g_B(\tilde{\tau};q_0=1/2)= - 
\left({{\pi}\over{\sin\pi \tilde{\tau}}}
\right)^{2\Delta_B}
\end{equation}
with the periodicity requirements 
$g_f(\tilde{\tau}+1)=-g_f(\tilde{\tau})$, 
$g_B(\tilde{\tau}+1)=g_B(\tilde{\tau})$. 
Note that these functions satisfy the additional symmetry 
$g_{f,B}(1-\tilde{\tau})=g_{f,B}(\tilde{\tau})$ indicating that 
{\it they can only apply 
to the particle-hole symmetric case} $q_0=1/2$. The corresponding form of the 
scaling functions associated with the spectral densities (\ref{scalrho})
reads, with $\tilde{\omega}\equiv \omega/T$ (after a calculation detailed in 
Appendix \ref{detailcalcul}):
\begin{eqnarray}
\label{DensScalSym}
\nonumber &\phi_f(\tilde{\omega},q_0=1/2)=
{{1}\over{\pi}}(2\pi)^{2\Delta_f-1}\cosh{{\tilde{\omega}}\over{2}}\,
{{\Gamma(\Delta_f+i{{\tilde{\omega}}\over{2\pi}})
\Gamma(\Delta_f-i{{\tilde{\omega}}\over{2\pi}})}
\over{\Gamma(2\Delta_f)}}\\
&\phi_B(\tilde{\omega},q_0=1/2)=
{{1}\over{\pi}}(2\pi)^{2\Delta_B-1}\sinh{{\tilde{\omega}}\over{2}}\,
{{\Gamma(\Delta_B+i{{\tilde{\omega}}\over{2\pi}})
\Gamma(\Delta_B-i{{\tilde{\omega}}\over{2\pi}})}
\over{\Gamma(2\Delta_B)}}
\end{eqnarray}
Note that the $\omega\rightarrow 0$ singularity of the $T=0$ case is now  
recovered for $\omega\gg T$:
\begin{equation}\label{AsympPhif}
\phi_{f,B}(\tilde{\omega},q_0=1/2) 
\mathop{\sim}_{\tilde\omega \rightarrow + \infty} 
{1 \over \Gamma (2\Delta_{f,B})}
\left({{1}\over{\tilde\omega}}\right)^{1-2\Delta_{f,B}}
\end{equation}
It is an interesting calculation, performed in detail in Appendix
 \ref{detailcalcul}, 
to check that indeed these scaling functions 
do solve the large-$N$ equations (\ref{sp},\ref{defsigma}) 
at finite 
temperature in the scaling regime. 
That NCA-like integral equations obey the finite-temperature scaling 
properties dictated by conformal invariance has not, 
to our knowledge, been pointed out in the previous litterature.

\subsection{General values of $q_0$}

\subsubsection{Spectral asymmetry}

Let us move to the general case of representations with $q_0\neq{1\over2}$ 
in which the particle-hole symmetry between pseudo-fermions is broken. 
The {\it exponent} of the power-law singularity in the $T=0$ spectral
 densities is not affected by 
this asymmetry. It does induce however {\it an asymmetry of the prefactors} 
associated with positive and negative frequencies as $\omega\rightarrow 0$. 
We introduce an angle $\theta$ to parametrize this asymmetry, defined such 
that:
\begin{equation}
\rho_{f}(\omega\rightarrow 0^+)\sim h(\gamma,\theta)
{{\sin (\pi\Delta_f+\theta)}\over{\omega^{1-2\Delta_f}}}
\,\,\,,\,\,\,
\rho_{f}(\omega\rightarrow 0^-)\sim    h(\gamma,\theta)
{{\sin (\pi\Delta_f-\theta)}\over{(-\omega)^{1-2\Delta_f}}}
\end{equation}
where $ h(\gamma,\theta)$ is a constant prefactor. 
The explicit dependence of $\theta$ on $q_0$ will be derived below. 
This corresponds to the following analytic behaviour of the Green's 
function in the complex frequency plane, as $z\rightarrow 0$:
\begin{equation} 
\label{singGf}
G^R_f(z) \sim  h(\gamma,\theta) {e^{-i\pi \Delta_f -i\theta}\over 
z^{1-2\Delta_f}} \,\,\,\, \Im z >0
\end{equation}
Equivalently, this means that the symmetry $G_f(\beta-\tau)=G_f(\tau)$ 
is broken, and that the scaling function $g_f(\tilde{\tau})$ must satisfy 
(from the behaviour of its Fourier transform):
\begin{equation}
\label{asym}
{{ g_f(0^+)}\over{ g_f(1^-)}} = 
{{ \sin (\pi\Delta_f+\theta)}\over {\sin (\pi\Delta_f-\theta)}}
\end{equation}
We have found, by an explicit analysis of the saddle-point and constraint 
equations in the scaling regime, which is detailed in 
Appendix \ref{detailcalcul}, 
that the full scaling functions for this 
asymmetric case are very simply related to the symmetric ones at 
$q_0=1/2$, through:
\begin{equation}
\label{ansatzCasGene}
g_{f,B}(\tilde{\tau};q_0)={{e^{\alpha(\tilde{\tau}-{{1}\over{2}})}}
\over{\ca}}
g_{f,B}(\tilde{\tau};q_0=1/2)\,=
{{e^{\alpha(\tilde{\tau}-{{1}\over{2}})}}\over{\ca}}\,
\left({{\pi}\over{\sin\pi \tilde{\tau}}}\right)^{2\Delta_{f,B}}
\end{equation} 
where the parameter $\alpha$ is simply related to $\theta$ so as to obey 
(\ref{asym}):
\begin{equation}
\label{RelationAlphaTheta}
\alpha =\ln {\sin
 \left( {\pi\over 2(1+\gamma)} - \theta \right) \over \sin 
\left( {{\pi \over 2(1+\gamma)}+\theta}  \right) }
\end{equation}
Fourier transforming, this leads to the scaling functions for the 
spectral densities:
\begin{eqnarray}\label{DensiteSpectraleGenerales}
\nonumber
\phi_f(\tilde{\omega},q_0)&=&
{{ \cosh {{\tilde{\omega}}\over{2}} }\over{\cosh {{\tilde{\omega}+\alpha}
\over{2}}\ca }}\,
\phi_f(\tilde{\omega}+\alpha,q_0=1/2)\\
\phi_B(\tilde{\omega},q_0)&=&{{ \sinh {{\tilde{\omega}}\over{2}} }
\over{\sinh {{\tilde{\omega}+\alpha}\over{2}}\ca }}\,
\phi_B(\tilde{\omega}+\alpha,q_0=1/2)
\end{eqnarray}

\begin{figure}
\fig{15cm}{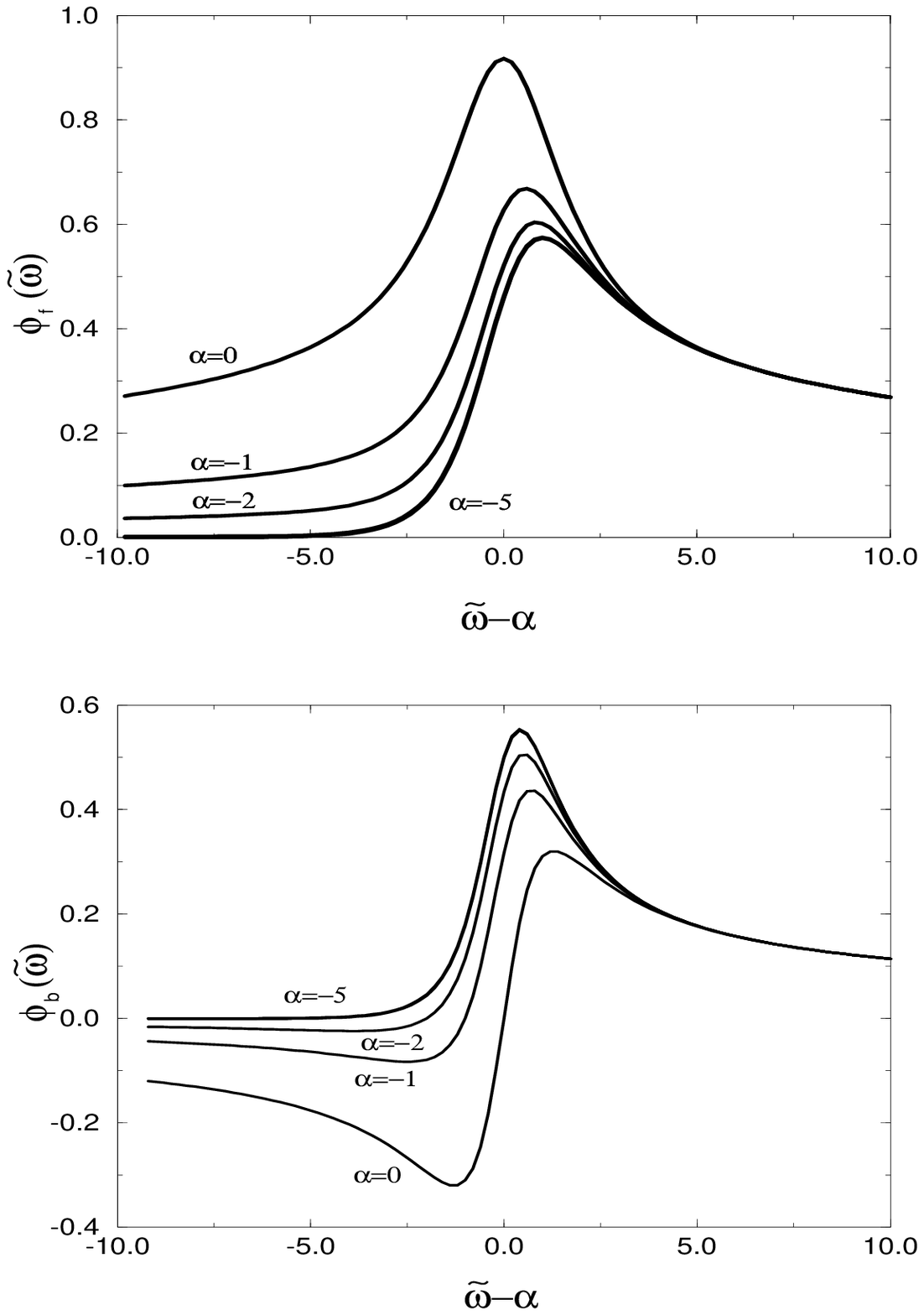} 
\vskip 1cm
\caption{ \label{Dos}
Plot of $ \phi_f$ and $ \phi_b$
as a function of $\tilde\omega -\alpha$ 
for different values of the asymmetry  parameter $\alpha$
 : $\alpha=0,-2,-5,-\infty$($\Delta_f=.3$)
}
\end{figure}

The thermal scaling function for the fermionic spectral density $\phi_f$ and
the bosonic one $\phi_{b}$
are plotted in Fig.(\ref{Dos}) 
 for various values of the asymmetry parameter
$\alpha$.  We also note the expression for the maximally asymmetric case
$\alpha \rightarrow -\infty$ (corresponding, as shown below, to
$q_0\rightarrow 0$, i.e. to the limit $Q\ll N$ as in the usual NCA).  
\begin{equation}
\phi_f(\tilde{\omega}-\alpha) \mathop{\sim}_{\alpha
  \rightarrow - \infty}  e^{\tilde\omega\over 2 } {(2 \pi)^{2\Delta_f-1}
  \Gamma\left( \Delta_f+i {\tilde{\omega} \over 2\pi} \right) 
 \Gamma\left(\Delta_f-i {\tilde{\omega}\over 2\pi}\right)
 \over \pi\Gamma(2\Delta_f)}
\end{equation}

We also note, for further use, 
the expressions of the full Green's functions 
in the complex frequency plane (defined by 
$g_{f/B}(z,\alpha )\equiv \int_{-\infty}^{+\infty} d\tilde\omega 
{{\phi_{f/B}(\tilde\omega)}\over{z-\tilde\omega}}$), in the 
scaling regime for $\Im z > 0 $:
\begin{eqnarray}
\label{fullGreenGenerales}
\nonumber
g_{f}(z,\alpha )&=& 
- {{2 i(2\pi)^{2\Delta_f-1}}\over{\ca \Gamma(2\Delta_f)\sin 2\pi \Delta_f}}
\times \\
&&\Gamma \left(\Delta_f+i{{z+\alpha }\over{2\pi}} \right)
\Gamma \left( \Delta_f-i{{z+\alpha}\over{2\pi}}\right)
\cos \left( \pi \Delta_{f}-{i\alpha  \over 2 } \right)
 \sin\left( \pi \Delta_{f}+i{z+\alpha \over 2} \right) \\ 
\nonumber
g_{B}(z,\alpha )&=&
- {{2 (2\pi)^{2\Delta_B-1}}\over{\ca \Gamma(2\Delta_B)\sin 2\pi \Delta_B}}
\times \\
&&\Gamma\left(\Delta_B+i{{z+\alpha }\over{2\pi}} \right)
\Gamma \left(\Delta_B-i{{z+\alpha}\over{2\pi}} \right)
\sin \left( \pi \Delta_{B}-{i\alpha  \over 2 }\right)
\sin \left( \pi \Delta_{B}+i{z+\alpha \over 2}\right) 
\end{eqnarray}

The reader interested in the details of these calculations is directed to 
Appendix \ref{detailcalcul}. 

At this stage, the point which remains to be clarified is the explicit 
relation between the asymmetry parameter $\theta$ and the parameter $q_0$ 
specifying the representation. This is the subject of the next section. 

Before turning to this point, we briefly comment on the CFT interpretation 
of the asymmetry parameter $\theta$ (or $\alpha$) associated with the 
particle-hole asymmetry of the fermionic fields. 
The form (\ref{cftscaling}) of the correlation functions at finite 
temperature 
in the scaling limit can be viewed as those of the exponential of a
compact bosonic field with periodic boundary conditions. The asymmetric
generalization (\ref{ansatzCasGene}) corresponds to a shifted boundary 
condition on the boson ({\it i.e} to a twisted boundary condition for 
its exponential). 

\subsubsection{Relation between $q_0$ and $\theta$}

Let us clarify the relation between the spectral asymmetry parameter 
$\theta$, and the parameter $q_0$ specifying the spin representation. 
That such a 
relation exists in universal form is a remarkable fact: indeed $\theta$ is 
{\it a low-energy parameter} associated with the low-frequency 
behaviour of the spectral density, 
while $q_0$ is the total pseudo-fermion 
number related by the constraint equation (\ref{eqlambda}) to an 
integral of the spectral density {\it over all frequencies}.  
The situation is similar to that of the Friedel sum rule in impurity models, 
or to Luttinger theorem in a Fermi liquid, and indeed 
the derivation of the relation between $q_0$ and $\theta$ follows 
similar lines \cite{AGD}. 
It is in a sense 
a Friedel sum rule for the quasiparticles carrying the spin degrees of 
freedom (namely, the pseudofermions $f_{\alpha}$).

We start from the constraint equation (\ref{eqlambda}) written at 
zero-temperature as:
\begin{equation}
\label{q01}
q_0= -i \lim_{t\rightarrow0^+} \int {d \omega\over 2\pi} G_f(\omega)
 e^{i\omega t}
\end{equation}
In this expression, and below in this section, $G_f(\omega)$ 
and $G_B(\omega)$  
denote the (Feynman) $T=0$ Green's functions 
while the retarded Green's functions are denoted  by $G^R$. 
Using analytic continuation of (\ref{defsigma}), we have :
\begin{eqnarray}
{\partial \ln G_f(\omega)\over \partial \omega} - 
G_f(\omega) {\partial \Sigma_f(\omega)\over \partial \omega} 
&=& -G_f(\omega) \\
{\partial \ln G_B(\omega)\over \partial \omega} - 
G_B(\omega) {\partial \Sigma_B(\omega)\over \partial \omega} &=& 0
\end{eqnarray}
so that (\ref{q01}) can be rewritten as: 
\begin{equation}
q_0= i \lim_{t\rightarrow0^+} \int {d \omega\over 2\pi} 
\left[ {\partial \ln G_f(\omega)\over \partial \omega} - G_f(\omega) 
{\partial \Sigma_f(\omega)\over \partial \omega}  - 
\gamma \left({\partial \ln G_B(\omega)\over \partial \omega} - 
G_B(\omega) {\partial \Sigma_B(\omega)\over \partial \omega}  \right)
 \right ] e^{i\omega t}
\end{equation}
In this expression, the bosonic part (which vanishes altogether) has been 
included in order to transform further the terms involving derivatives of 
the self energy, using analyticity. This transformation is only possible 
if both fermionic and bosonic terms are considered. This is because the 
Luttinger-Ward functional \cite{AGD} of this model involves both Green's 
functions. It has a simple explicit expression which reads:
\begin{equation}
\label{Lutt-Ward}
\Phi_{LW}(G_{f,\alpha},G_{B,i}) = 
\sum_{\alpha,i} \int d t G_0(t) G_{f,\alpha}(-t) G_{B,i}(t)
\end{equation}
such that the saddle-point equations (\ref{sp}) are recovered by derivation:
\begin{equation}
\Sigma_{f,\alpha}(t) = {\delta \Phi_{LW}\over \delta G_{f,\alpha}(-t)}
\,\,\,,\,\,\,
\Sigma_{B,i}(t) = -{\delta \Phi_{LW}\over \delta G_{B,i}(-t)}
\end{equation}

\null From the existence of $\Phi_{LW}$, we obtain the sum rule:
\begin{equation}
\int_{-\infty}^{\infty} d \omega\left( 
\Sigma_f(\omega) {\partial G_f(\omega)\over \partial
\omega} - \gamma \Sigma_B(\omega) {\partial G_B(\omega)\over \partial \omega}
 \right) =0 
\end{equation}
(Note that there is no logarithmic divergence in this integral).
After integrating by parts and using the asymptotic behaviour of the two 
Green's functions in order to eliminate the boundary terms, we get : 
\begin{equation}
\label{q02}
q_0 = i \int_{-\infty}^{\infty} {d \omega\over 2\pi} 
\left({\partial \ln G_f(\omega) \over \partial \omega} - 
\gamma {\partial \ln G_B(\omega) \over \partial \omega} \right)e^{i\omega0^+}
\end{equation}
Since $G(\omega)=G^R(\omega)$ for $\omega>0$ and $G(\omega)=
\overline{G^R(\omega)}$ 
for $\omega<0$ (with $G^R$ the retarded Green's function), this can be 
transformed using:.
\begin{equation}
\int_{-\infty}^{\infty} {d \omega\over 2\pi} 
{\partial \ln G_{f,B}(\omega) \over \partial \omega} e^{i\omega0^+} = 
\int_{-\infty}^{\infty} {d \omega\over 2\pi} {\partial \ln G^R_{f,B}(\omega) 
\over \partial \omega} e^{i\omega0^+} + 
\int_{-\infty}^{0} {d \omega\over 2\pi}{\partial 
\over \partial \omega} \ln \left( \overline{G^R_{f,B}(\omega)}
\over G^R_{f,B}(\omega) \right)  e^{i\omega0^+}
\end{equation}
The first integrals in the right hand side 
can be deformed in the upper plane and their sum vanishes \cite{AGD}. 
Thus we obtain (denoting by $\arg G$ the argument of $G$) :
\begin{equation}
\label{Formuleq0}
\pi q_0 = \arg G^R_f(0^-) - \arg G^R_f(-\infty) - 
\gamma \left( \arg G^R_B(0^-)- \arg G^R_B(-\infty)\right)
\end{equation}
$\arg G^R_f(0^-)$ directly follows from the  parametrisation (\ref{singGf}) 
defining $\theta$. It can also be read off from the behaviour of the 
scaling function $g_f(z)$ for $z=\pm\infty$. Thus, from 
(\ref{fullGreenGenerales}) we can also read off $\arg G^R_B(0^-)$: 
\begin{equation}
\arg G^R_f(0^-) = \pi\Delta_f  - \theta - \pi
\,\,\,,\,\,\,
\arg G^R_B(0^-) = \theta - \pi \Delta_f
\end{equation}
Taking into account that 
$G^R_f(\omega) \mathop{\sim}_{\omega\rightarrow -\infty} 1/\omega$ 
and that $\Im G^R_f <0$ we have $\arg G^R_f(-\infty) = -\pi$. 
Similarly, we have $\arg G^R_B(-\infty)=0$. 
Inserting these expressions into 
(\ref{Formuleq0}), we finally obtain the desired relation between 
$q_0$ and $\theta$ (or $\alpha$): 
\begin{equation}
\label{RelationQ0Theta}
{\theta (1+\gamma)\over \pi} = {1\over 2} - q_0\,\,\,,\,\,\,
\alpha=\ln {{ \sin {{\pi q_0}\over{1+\gamma}} }\over{
\sin {{\pi (1-q_0)}\over{1+\gamma}} }}
\end{equation}
This, together with (\ref{fullGreenGenerales}), fully determines the 
universal scaling form of the spectral functions in the low-frequency, 
low temperature limit.

\section{Physical quantities and comparison with the CFT approach}

\subsection{Impurity residual entropy at $T=0$}

The impurity contribution to the free-energy (per colour of spin) 
$f_{imp}=(F-F_{bulk})/N$ 
reads, at the saddle point:
\begin{equation}
\label{free1}
 f_{imp} =  q_0 \lambda +
\,T\sum_n \mbox{ln} G_f(i\omega_n) - 
\gamma\, T\sum_n \mbox{ln} G_B(i\nu_n) - \int_{0}^{\beta}d \tau 
\Sigma_{f}(\tau)G_{f}(-\tau )
\end{equation}
This expression can be derived either directly from the saddle-point 
effective action (in which case the  
last term arises from the quadratic term in $Q$ and $\overline{Q}$), or 
from the relation between the free-energy and the Luttinger-Ward 
functional. 
$F_{bulk}= N^{2}\gamma  T \mbox{Tr} \ln G_{0} $ 
is the free energy of the conduction electrons.
In (\ref{free1}) the formulas $\mbox{Tr} \ln G$ are ambiguous.
 We must precisely 
define which regularisation of these sums we consider :
 the actual value of the sums depends on the 
precise definition  of the functional integral.
For the fermionic field, the standard procedure of  
adding and substracting the contribution of a free local fermion, and
introducing an oscillating term to regularize the Matsubara sum holds:
\begin{equation}\label{regulfermions}
\mbox{Tr} \ln G_{f}= - T \ln  2 + T\sum_{n} \ln 
\left( i\omega_{n}G_{f}(i\omega_{n}) \right)
e^{i\omega_{n}0^{+}}
\end{equation}
The situation is somewhat less familiar for the bosonic field. As 
pointed out above, the latter is merely a commuting auxiliary field 
(rather than a true boson). We have found that the correct regularisation 
to be used is:
\begin{equation}\label{regulbosons}
\mbox{Tr} \ln G_{B}= T\lim_{N\rightarrow \infty }\sum_{n=-N}^{n=N}
 \ln \left({J G_{B}(i\omega_{n})} \right)
\end{equation}
The factor of $J$ takes into account the determinant                           
introduced by the decoupling with $B$, and a {\it symmetric} definition 
of the (convergent) Matsubara sum has been used.
Some  details and justifications about these regularisations are given in 
Appendix (\ref{calculentropy}).

We shall perform a low-temperature expansion of Eq.(\ref{free1}), considering 
successively the particle-hole symmetric ($q_0=1/2$) and asymmetric  
($q_0\neq 1/2$) cases, which require rather different treatments.  

\subsubsection{The particle-hole symmetric point $q_{0}=\frac{1}{2}$}

In this case $\lambda =0$, so that the first term in Eq.(\ref{free1}) does 
not contribute. Let us consider the last term in (\ref{free1}). 
Using the spectral representation of $G_{f}$ and
  the definition of $\Sigma_{f}$ we obtain easily (for $\lambda=0$) :
\begin{equation}
\Psi \equiv  \int_{0}^{\beta}d \tau \Sigma_{f}(\tau)G_{f}(-\tau )=
 \int_{-\infty}^{+\infty} d \omega \,\, \frac{\omega \rho_{f}(\omega )}
{1+e^{\beta \omega }}  
\end{equation}
We substract the value at $T=0$ :
\begin{equation}\label{psi}
\Psi(T)- \Psi(T=0) =
-\int_{0}^{\infty} d \omega \,\, \omega 
\left(\rho_{f}(\omega,T)-\rho_{f}(\omega,T=0) \right )
+ 2 \int_{0}^{\infty}d \omega \,\, \frac{\omega \rho_{f}(\omega )}
{1+e^{\beta \omega }}  
\end{equation}
In the second term, we can replace $\rho_{f}$ by its scaling limit.
So this term is of order $O(T^{2\Delta_{f}+1})$.
We know the asymptotics of $\phi_{f}$ : $\phi_{f}(x)\sim_{x\rightarrow \infty}
C_{1} x^{2\Delta_{f}-1} + C_{2} x^{2\Delta_{f}-3}$. (The term
 $x^{2\Delta_{f}-2}$ cancels due to the particle-hole symmetry).
Thus, the first term in (\ref{psi}) is of the form :
$T^{2\Delta_{f}+1} \int_{0}^{\infty} dx \,\, x(\phi_{f}(x)-
C_{1}x^{2\Delta_{f}-1})$ (the integral is convergent). 
We conclude that  $\Psi (T) = \Psi (0) + O(T^{2\Delta_{f}+1})$, so 
that the last term in (\ref{free1}) does not contribute to the 
zero-temperature entropy in the particle-hole symmetric case.

Let us express the remaining terms in (\ref{free1}) 
as integrals over real frequencies, using the 
regularisations introduced above. 
As detailed in Appendix \ref{calculentropy}, this leads to the 
following expression, involving the
argument of the (finite-temperature) retarded Green's functions:
\begin{equation}\label{free}
  f_{imp} = 
{{1}\over{\pi}} \int_{-\infty}^{+\infty} d\omega \,
\left[n_F(\omega) 
\left(\arctan \left(G'_f(\omega)\over G''_f(\omega) \right) - 
{{\pi}\over{2}}  \right)
 -\gamma n_B(\omega) \arctan \left(G''_B(\omega)\over G'_B(\omega) \right)
 \right] 
\end{equation}
In these expressions $n_F$ (resp. $n_B$) are the Fermi (resp. Bose) factor.

At this point, it would seem that in order to perform a low-temperature 
expansion of the free-energy, one has to make a Sommerfeld expansion of the 
Fermi and Bose factors. This is not the case however, for two reasons: 
i) the argument of the Green's functions appearing in (\ref{free}) are 
{\it not} continuous at $\omega=0$, so that a linear term in $T$ does 
appear (as expected from the non-zero value of $S_{imp}$) ii) 
the Green's functions have an {\it intrinsic} temperature dependence, 
and the full scaling functions computed above must be used in 
(\ref{free}). 
More precisely, when computing  the difference $f_{imp}(T)-f_{imp}(T=0)$,
the leading term is obtained by replacing the Green's function by their 
scaling form (\ref{fullGreenGenerales}).

These considerations lead to the following expression of the impurity 
entropy (per spin colour) $s_{imp}=S_{imp}/N$ 
at zero-temperature for $q_{0}=\frac{1}{2}$ :
\begin{eqnarray}
\label{Entropy}
\nonumber s_{imp} &=& - {{1}\over{\pi}} \int_{-\infty}^{0} d{\tilde{\omega}} 
\left( a_f({\tilde{\omega}})-a_f({\tilde{\omega}}=-\infty) \right) - 
{{1}\over{\pi}} \int_{-\infty}^{+\infty} d{\tilde{\omega}} 
{{1}\over{e^{|{\tilde{\omega}}|}+1}} a_f({\tilde{\omega}}) 
\mbox{sgn}({\tilde{\omega}})   \\
&& -{{\gamma}\over{\pi}} \int_{-\infty}^{0} d{\tilde{\omega}} 
\left( a_b({\tilde{\omega}})-a_b({\tilde{\omega}}=-\infty) \right)
+ {{\gamma}\over{\pi}} \int_{-\infty}^{+\infty} d{\tilde{\omega}} 
 a_b({\tilde{\omega}}) \mbox{sgn}({\tilde{\omega}}) 
{{1}\over{e^{|{\tilde{\omega}}|}-1}}
\end{eqnarray}

In this expression, $a_{f,B}$ denotes the arguments of the scaling
functions, obtained from (\ref{fullGreenGenerales}): 
\begin{eqnarray}
a_f({\tilde{\omega}}) &\equiv & \arctan {{g'_f({\tilde{\omega}})}
\over{g''_f({\tilde{\omega}})}} = 
- \arctan \left( \cot(\pi\Delta_f)\tanh \frac{{\tilde{\omega}} }{2} \right) \\
a_B({\tilde{\omega}})&\equiv&  \arctan {{g''_B({\tilde{\omega}})}
\over{g'_B({\tilde{\omega}})}} = 
\arctan \left( \tan(\pi\Delta_f)\tanh \frac{{\tilde{\omega}} }{2} \right)
\end{eqnarray}

 From (\ref{Entropy}), we obtain with 
$t= \tan {\pi \Delta_{f}}$
\begin{equation}
{ s_{imp}\over \gamma+1} = -{2\over \pi} \int_0^1 d u \,
\left(  {2 \arctan t\over \pi (1-u^2)} \left(
 u \arctan \left({u\over t}\right) + {\arctan(u t)\over u }\right) -
 {\arctan(u t)\over u(1-u^2)}\right)
\end{equation}
To perform the integration, we note that  
$ {\partial \over \partial t } 
\left(  {\partial \over \partial t} 
\left((1+t^2) {s_{imp} \over 1+\gamma  }\right)\right)
 = -{2 t\over \pi (1+t^2)} $
and we  obtain finally the simple expression 
\begin{equation}
\label{entropyPH}
 s_{imp}(q_{0}=\frac{1}{2}) =
 \ln 2 - {\gamma+1\over \pi} \int_0^{\tan{\pi\over 2(1+\gamma)}}
 {\ln(1+u^2)\over(1+u^2)} du
\end{equation}
This can also be rewritten, after a change of integration variable, as:
\begin{equation}
 s_{imp}(q_0=1/2)\equiv {{1}\over{N}} S_{imp} =
 {1+\gamma\over \pi} \left[  f \left( {\pi\over 1+\gamma}\right)
- 2 f \left( {\pi\over 2(1+\gamma)} \right) \right]
\end{equation}
with
$$
f(x) \equiv \int_0^x \ln\sin (u) \,\, d u
$$
This coincides with the large-N limit of the CFT result, 
Eq.(\ref{entropieImpureteCFT}), in the particle-hole symmetric case.

\subsubsection{The general case $q_0\neq 1/2$}

\begin{figure}
\[
\fig{10cm}{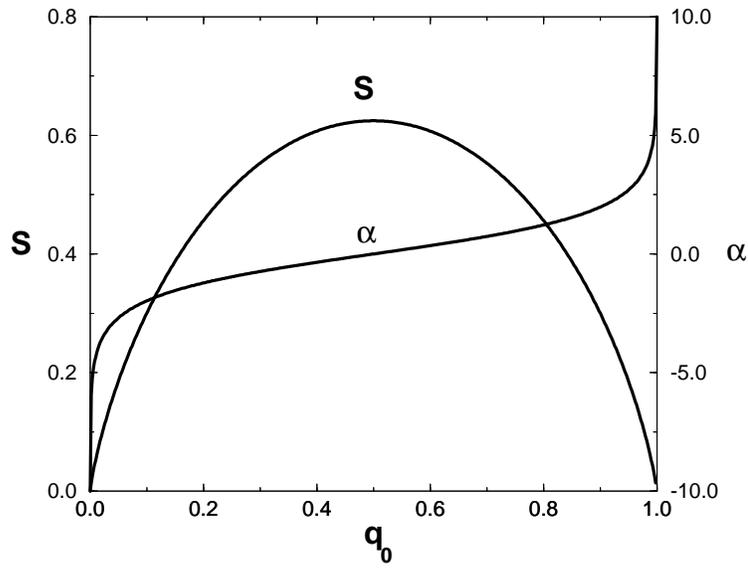}
\] 
\vskip 1cm
\caption{ \label{FigEntropie}
Residual entropy $s_{imp}$ and $\alpha $ vs. $q_{0}$ for $\gamma =1.5$} 
\end{figure}

For $q_{0}\neq \frac{1}{2}$, the first term in Eq.(\ref{free1}) also 
contributes to the entropy. Indeed, as shown below, the Lagrange multiplier 
$\lambda(T)$ at the saddle-point has a term which is linear in temperature. 
This stems from a very general thermodynamic relation, which is derived by 
taking the derivative of $Z$ with respect to $q_0$ in 
the functional integral, leading to:
\begin{equation}  
\left\langle  -\frac{1}{\beta }\int_{0}^{\beta } i\mu (\tau )  \right\rangle
={\partial F \over \partial q_{0} }
\end{equation}  
where the average is to be understood with 
the action (\ref{action1}). At the saddle-point, we thus have :
$\lambda = {\partial f_{imp} \over \partial q_{0} }$ and in particular :
\begin{equation}\label{RelationLambdaS}
\left. {\partial \lambda  \over \partial T }\right|_{T=0}=
  -{\partial s_{imp} \over\partial q_{0} }
\end{equation}
We shall directly use this equation in order 
to compute the residual entropy, by calculating the  
linear correction in $T$ to $\lambda$, and then integrating over $q_0$. 
This method shortcuts the  
full low-temperature expansion of the free-energy (as done in the previous 
section), which actually turns out to be quite a difficult task to perform 
correctly for $q_0\neq 1/2$ \cite{footentro}. 
In order to calculate this linear correction, we shall   
relate $\lambda $ to the {\it high-frequency behaviour} of the fermion 
Green's function. 
As $\Sigma_{f}(i\omega_{n} )\rightarrow 0$ when $\omega_{n} \rightarrow
 \pm \infty $  we have :
\begin{equation}
G_{f} (i\omega_{n}) = \frac{1}{i\omega_{n} }- 
\frac{\lambda }{(i\omega_{n} )^{2}} +
o\left(\frac{1}{(i\omega_{n} )^{2}}\right)
\end{equation}
This shows that $-\lambda $ is the discontinuity of 
the derivative of $G_{f}(\tau)$ with respect to $\tau$ at $\tau=0$:
\begin{equation}\label{lambdadisc}
\partial_{\tau }G_{f}(0^{+}) +
 \partial_{\tau }G_{f}(\beta^{-}) = 
\int_{-\infty}^{+\infty} d\omega \omega\rho_f(\omega) = -\lambda
\end{equation}
Let us define $g(\tau)$ by:
\begin{equation}
G_{f}(\tau ) = 
{{e^{\alpha (\frac{\tau}{\beta}  -\frac{1}{2})}}\over{\ca}}\,g(\tau)
\label{defnewg}
\end{equation}
where $\alpha$ is the spectral asymmetry parameter in 
Eq.(\ref{RelationQ0Theta}) (at this stage we emphasize  that the 
full finite temperature, finite cutoff, Green's function $G_f$ is considered). 
Equation (\ref{lambdadisc}) can be rewritten as:
\begin{equation}
\lambda =\alpha T +  \tanh \left(\frac{\alpha }{2} \right)
\left(\partial_{\tau }g(0^{+}) - \partial_{\tau }g(\beta ^{-})  \right)
- \left(\partial_{\tau }g(0^{+}) + \partial_{\tau }g(\beta ^{-})  \right)
\end{equation}
where we have used that $G_f(0^+)+G_f(\beta^-)=-1$. 
Denoting by $\rho_g(\omega)$ the spectral function associated with $g$, 
we have: 
\begin{equation}
\partial_{\tau }g(0^{+}) - \partial_{\tau }g(\beta ^{-})
= \int d \omega  {\omega 
\left( \rho_{g}(\omega)-\rho_{g}(-\omega ) \right)  
  \over
1 + e^{-\beta \omega }}
\label{extraterm}
\end{equation}
In the scaling limit, the spectral function $\rho_g$ must become 
{\it particle-hole symmetric} (since the effect of the 
particle-hole asymmetry in this limit is entirely captured by $\alpha$ in 
(\ref{defnewg})), and must coincide with 
$A_fT^{2\Delta_f-1}\phi_f(\omega/T;q_0=1/2)$. Hence, following the same 
reasoning than for $\Psi$ above,  
the term in (\ref{extraterm}) is of order 
$\mbox{const.}+O(T^{2\Delta_{f}+1})$. 
Thus we have :
\begin{equation}\label{res1}
\left. {\partial \lambda  \over \partial T }\right|_{T=0}= \alpha 
- \left. {\partial A  \over \partial T }\right|_{T=0} 
\end{equation}
where
 $A=\partial_{\tau }g(0^{+}) + \partial_{\tau }g(\beta ^{-})$ is 
the discontinuity of the derivative $\partial_{\tau}g$.
$A$ reflects the particle-hole asymmetry of $g$ and thus vanishes in the 
scaling limit. Actually the derivative $\frac{\partial A}{\partial T}$
also vanishes as $T\rightarrow 0$ as we now show.
Consider first sending the bare cutoff to infinity (along with $J$)
so as to keep the Kondo temperature fixed. 
In this limit $A$ takes the form: 
 $A= T f\left(\frac{T}{T_{K} } \right)$.
The low energy scaling limit, in which Eq. (\ref{ansatzCasGene}) holds, 
can be reached by fixing $T$ and sending $T_{K}$ to infinity.
Since $g$ must become particle-hole symmetric in this limit,
 this implies  that $f(x)$ vanishes at small $x$. Hence, taking a derivative
 with respect to  temperature, of $A= T f\left(\frac{T}{T_{K} } \right)$ 
we find that $\left.\frac{\partial A}{\partial T}\right |_{T=0}=0$.
Thus we finally obtain:
\begin{equation}\label{OUF}
{\partial s_{imp} \over \partial q_{0} } = 
- \left.{\partial \lambda  \over \partial T }\right|_{T=0} = - \alpha 
\end{equation}
where $\alpha(q_0)$ is given in Eq.(\ref{RelationQ0Theta}). 
Integrating this equation over $q_0$ (taking into account as a boundary 
condition the value of $s_{imp}(q_0=1/2)$ obtained above), we finally 
derive the expression of the entropy: 
\begin{equation}
\label{EntropyCFT}
 s_{imp}\equiv{{1}\over{N}} S_{imp}=
 {1+\gamma\over \pi} \left[  f \left({\pi\over 1+\gamma}\right)
- f \left({\pi\over 1+\gamma}(1-q_0)\right) - f \left({\pi\over 1+\gamma}
 q_0\right) \right] 
\end{equation}
with, as above: $ f(x) \equiv \int_0^x \ln\sin (u) \,\, d u $. 
The expression (\ref{EntropyCFT}) 
coincides precisely with the large N limit of the CFT result 
(\ref{entropieImpureteCFT})  \cite{KC}.
A plot of the residual entropy and of the asymmetry parameter $\alpha$ as 
a function of $q_0$ is displayed on Fig.\ref{FigEntropie}. 
$s_{imp}$ is maximal at $q_0=1/2$ and vanishes as $q_0\rightarrow 0$ as 
expected. 

In this section, we have discovered that the spectral asymmetry parameter 
(``twist'') $\alpha$ shares a fairly simple relation with the residual 
entropy, given by Eq.(\ref{OUF}). These are two universal quantities, 
characteristic of the fixed-point. Remarkably, $\alpha$ also coincides 
with the term proportional to $T$ in $\lambda$ (while $\lambda$ itself 
is non-universal, its linear term in $T$ is). 
It is tempting to speculate that a deeper interpretation of these 
facts is still to be found. 

\subsection{Internal Energy and Specific Heat}

The low-temperature behaviour of the internal energy in the large-$N$ limit 
can be obtained by two different methods. We shall briefly describe both 
since they emphasize different and complementary aspects of the physics.

In the first method, we use the effective action in the form (\ref{action1}), 
{\it before} the decoupling with the auxiliary bosonic field $B_i(\tau)$ is 
made. We thus have a quartic interaction vertex between the conduction 
electrons at the origin and the Abrikosov fermions representing the 
quasiparticles in the spin sector, which reads:
\begin{equation}
\label{kondoint}
{J \over N} \sum_{1\leq\alpha,\beta\leq N}
(f^{\dagger}_{\beta}f_{\alpha}-{Q \over N}\delta_{\alpha\beta})
\sum_{i=1}^K c^{\dagger}_{i\alpha}c_{i\beta}
\end{equation}
One can then perform a skeleton expansion of the free-energy functional in 
terms of the {\it interacting} Green's functions for the pseudo-fermions and 
the conduction electrons, $G_f(\tau)$ and $G_c(\tau)$. 
The first-order (Hartree) contribution to this functional vanishes 
because the spin operator in (\ref{kondoint}) is written in a traceless 
manner. The next contribution, at second order, yields the most singular 
contribution at low temperature and reads:
\begin{equation}
\label{energ2opt}
\Delta E \propto J^2\,\int_{0}^{\beta} d\tau G_c(\tau)^2 G_f(-\tau)^2
\end{equation}
At the saddle-point, the interacting conduction electron Green's 
function is $G_c(\tau)\propto G_f(\tau)G_B(-\tau)$, and hence its 
dominant long-time behaviour is: $G_c(\tau)\sim 1/\tau$. Inserting 
this, together with $G_f(\tau)\sim 1/\tau^{2\Delta_f}$ in 
(\ref{energ2opt}), we see that the leading low-temperature behaviour to 
the energy reads $\Delta E \propto c_1 T^{4\Delta_f+1} +c_2 T^2 +\cdots$, 
and hence to the specific heat coefficient: 
\begin{eqnarray}
\label{specific}
\nonumber &
\gamma>1\,\,\,:\,\,\, 
C/T \sim T^{4\Delta_f-1} \sim \left( {1\over T}\right)
^{\gamma-1 \over \gamma+1} \\
\nonumber & \gamma=1\,\,\,:\,\,\, C/T \sim  \ln 1/T\\
& \gamma<1\,\,\,:\,\,\, C/T\sim \mbox{const.}
\end{eqnarray} 
which agrees with the CFT result described above. We note that there is a 
quite precise connection between this calculation and the CFT approach: 
the operator appearing in the Kondo interaction (\ref{kondoint}) acquires 
conformal dimension $2(\Delta_c+\Delta_f)=1+2\Delta_f$ and has the 
appropriate structure of the scalar product of a spin current with the 
operator $S_{\alpha\beta}$ (transforming as the adjoint). Therefore, it is 
the large-$N$ version of the leading irrelevant operator associated with 
the spin sector, as described in Sec.\ref{CFTsec}. It is satisfying that 
the leading low-$T$ behaviour comes from the second-order contribution 
of this operator in this formalism as well. 
We note that for the simple 
Kondo model (\ref{ham1}), in the scaling limit, 
the analogous irrelevant operator in the flavour sector  
{\it does not} show up in the calculation of the energy in the 
large-N solution. We shall comment further on this point below. 

The second method to investigate the internal energy is to 
push the low-temperature expansion of the free-energy to higher orders. 
To this end, we need to compute higher-order terms in the expansion 
(\ref{scalg}) of the Green's functions in the scaling regime. 
This computation is detailed in Appendix \ref{detailcalcul}.3, and leads to:
\begin{eqnarray}
\label{defsndfntechelle}
G_{f}(\tau)&=& A_{f} \beta^{-2\Delta_f} g_f\left({\tau\over\beta}\right) + 
\beta^{-4\Delta_f} g_f^{(2)}\left({\tau\over\beta}\right) + 
\beta^{-6\Delta_f} g_f^{(3)}\left({\tau\over\beta}\right) + \cdots 
 \\
G_{B}(\tau)&=& A_B \beta^{-2\Delta_B} g_B\left({\tau\over\beta}\right) + 
\beta^{-1} g_B^{(2)}\left({\tau\over\beta}\right) + 
\beta^{-1-2\Delta_f} g_B^{(3)}\left({\tau\over\beta}\right) + \cdots 
\end{eqnarray}
Let us emphasize that the exponents appearing in this expansion 
{\it are not 
symmetric between the bosonic and fermionic degrees of freedom}. 
This is 
because we are dealing with the Kondo model for which the auxiliary  
field (bosonic) propagator has no frequency dependence 
in the non-interacting theory. Also, the expansion given in 
(\ref{defsndfntechelle}) assumes a perfectly flat conduction band in the limit 
of an infinite bandwith (conformal limit). 
Using this expansion into the expression (\ref{free}) of the free-energy 
leads to a specific heat coefficient: 
$C/T\sim c_0 T^{2\Delta_f-1} + c_1 T^{4\Delta_f-1} + c_2 +\cdots$. 
The coefficient $c_0$ actually vanishes, so that the behavior in 
(\ref{specific}) is recovered. The vanishing of $c_0$ was clear 
in the first approach, 
where it followed from the absence of Hartree terms. In the CFT approach, 
it is associated with the fact that the leading irrelevant operator 
does not contribute to the free energy at first order. The vanishing of 
$c_0$ implies non trivial sum rules relating the scaling functions 
$g_{f,B}$ and $g_{f,B}^{(2)}$ (which we have not attempted to check 
explicitly). 

We also note that this behaviour of the specific heat is modified when an 
Anderson model version of the present model is considered 
(as in Ref.\cite{CR}). Because the non-interacting slave boson 
propagator has a frequency dependence, the exponents of the 
second-order terms as written in (\ref{defsndfntechelle}) are only correct 
for $\gamma>1$ for the Anderson model. For $\gamma<1$, the term 
$\beta^{-4\Delta_f} g_f^{(2)}(\tau / \beta )$ is replaced by 
$\beta^{-1} g_f^{(2)}(\tau / \beta)$, while $\beta^{-1} 
g_B^{(2)}(\tau / \beta)$ 
is replaced by $\beta^{-4\Delta_b} g_B^{(2)}(\tau / \beta)$. 
As a result, one finds a diverging specific heat coefficient 
{\it in both cases}, with  
$C/T\sim T^{-(\gamma-1)/(\gamma+1)}$ for $\gamma>1$ and 
$C/T\sim T^{-(1-\gamma)/(\gamma+1)}$ for $\gamma<1$. The behaviour 
for $\gamma<1$ is due to the leading irrelevant operator in the flavor 
sector. Similarly, for $\gamma<1$ in the Anderson model, 
the susceptibility associated with 
the {\it flavour (channel) sector} $\chi_f$ is found to diverge \cite{CR}, 
so that a finite Wilson ratio can still be defined as $T\chi_f/C$ for 
$\gamma<1$. 

\subsection{Resistivity and T-matrix}
\label{resistsec}

In order to discuss transport properties, we define a scattering 
T-matrix for the conduction electrons in the usual manner (for a single 
impurity):
\begin{equation}
\label{defT}
G(\vec{k},\vec{k'},\omega+i0^+) = 
G_0(\vec{k},\omega+i0^+)\delta_{\vec{k},\vec{k'}} +  
G_0(\vec{k},\omega+i0^+) T(\omega) G_0(\vec{k'},\omega+i0^+)
\end{equation}
where $G$ and $G_0$ denote the interacting and 
non-interacting conduction-electron Green's functions, respectively. 
Taking a flat particle-hole symmetric band for the conduction electron and 
denoting by $\rho_0$ the local non-interacting density of states, this yields 
the local conduction electron Green's function in the form:

\begin{equation}
G(\omega+i0^+)\equiv 
\sum_{\vec{k},\vec{k'}} G(\vec{k},\vec{k'}) 
= -i\pi\rho_0 \left( 1-i\pi\rho_0 T(\omega) \right) 
\end{equation}
Following Ref.\cite{AL2}, we parametrize the zero-frequency limit 
of the T-matrix in terms of a scattering amplitude $S^1$ as:
\begin{equation}
T(\omega=0) \equiv - {{i}\over{2\pi\rho_0}} \, (1-S^1)
\end{equation}
so that, the zero frequency electron Green's function reads:
\begin{equation}
G(i0^+) = -i\pi\rho_0 {{1+S^1}\over{2}}
\end{equation}
$S^1=1$ corresponds to no scattering at all, while $S^1=-1$ corresponds 
to maximal unitary scattering ({\it i.e} $\pi/2$ phase shift and vanishing 
conduction electron density of states at the impurity site). 
In the overscreened case, as noted in \cite{AL2}, $S^1$ is in general such 
that $|S^1|<1$, reflecting the non-Fermi liquid nature of the model, and the 
fact that the actual quasiparticles bear no resemblance to the original 
electrons. In addition here, we shall find a new feature: namely that 
$S^1$ is in fact a {\it complex number} for non particle-hole 
symmetric spin representations ({\it i.e} $q_0\neq 1/2$).

We first derive an expression for $S^1$ for arbitrary $N$, $K$ and 
spin representation $Q=Nq_0$ by generalizing to $SU(N)$ the CFT 
approach of Ref.\cite{AL2}. There, it was shown that      
$S^{1}$ can be expressed as a ratio of elements of the modular S-matrix 
 $S_{\alpha, \beta}$  
of the $\widehat{SU}(N)_K$ algebra. Denoting by $0$ the identity 
representation, by $F$ the fundamental representation (corresponding to 
a Young tableau with a single box) and by $R$ 
the representation in which the impurity lives (Young tableau with a 
single column of $Q$ boxes), one has \cite{AL,AL2}:
\begin{equation}
\label{conformal1}
S^1\,=\,
{{S_{F,R}/S_{0,R}}\over{ S_{F,0}/S_{0,0} }}
\end{equation}
The evaluation of these elements of the modular S-matrix can be done along the
same lines than the
conformal field theory  calculation of the entropy, described above.
Some details are given in Appendix \ref{modsmatrix}.
The  result  is:
\begin{equation}
\label{conformal}
S^{1} = {\sin \left( \frac{(N+1) \pi}{N+K} \right) 
\exp \left(-i \frac{ \pi (1-2q_0)}{N+K} \right)
- \sin \left( {\pi  \over N+K } \right) \exp\left(-i{  \pi (N+1)(1-2q_0)\over
N+K } \right)
\over \sin\left({\pi N \over N+K } \right)}
\end{equation} 
Notice  that $S^{1}$ has both real and imaginary parts in
the absence of particle hole symmetry $q_{0}\neq\frac{1}{2}$.

Let us take the large-N limit of this expression, with $K/N=\gamma$ 
fixed. This reads, to first non-trivial order:
\begin{eqnarray}
\label{largeNS}
S^{1} &=& 1  + \frac{\pi}{N(1+ \gamma)} \left[ \cot \frac{\pi}{1+\gamma} -
 \frac{\cos \left( {\pi (1-2q_{0}) \over 1+\gamma } \right)}
{\sin \left(\frac{\pi}{1+\gamma} \right)} 
\right]
-  \frac{i\pi}{N(1+\gamma)}
\left[ 1-2q_0 -\frac{\sin \left({\pi (1-2q_0) \over 1+\gamma }\right)}
{\sin \left(\frac{\pi} {1+\gamma} \right)}
\right]
\end{eqnarray}
We now show how to recover this expression from an analysis of the 
 integral equations of the direct large-N solution. 
Coupling an external source to the conduction electrons in the functional 
integral formulation of the model, it is easily seen that the 
conduction electron T-matrix is given, in the large-N limit, by: 
\begin{equation}
T(\omega)\,=\, {1\over N} {\cal G}(\omega+i0^+)
\end{equation}
where ${\cal G}$ denotes the convolution of the fermion and auxiliary 
boson Green's function:
\begin{equation}
{\cal G}(\tau) =  G_{f}(\tau)G_{B}(-\tau) 
\end{equation}
Hence, we have at $T=0$ :
\begin{equation}\label{S1vsG}
S^{1} = 1 - \frac{2i \pi \rho_{0} }{N}  {\cal G} (i 0^{+})
\end{equation}

We first make use of the scaling limit of the two Green's functions,
 given by (\ref{ansatzCasGene}), and obtain the scaling form of 
${\cal G}$ : 
\begin{equation}
\label{gimpscal}
{\cal G}(\tau) = G_{f} (\tau) G_{B}(\beta-\tau) = 
\frac{\pi A_{f} A_{B}}{\beta\cb \sin {\pi \tau\over \beta}} + \cdots 
\end{equation}
We note that, {\it in this scaling limit}, the particle-hole 
asymmetry of the impurity  Green's function has been lost altogether 
: $\alpha$ has cancelled
completely in the $\tau$ dependence of $\cal G$ in this limit  
and is only present in the prefactor. Thus, only 
$\Re S^{1}$ can be extracted from the scaling limit, while 
$\Im S^{1}$ requires a more sophisticated analysis.  
Eq. (\ref{gimpscal}) implies $\Im {\cal G}=A_{f}A_{B}\pi$, and hence 
$\Re S^{1}-1  =  2 \pi^{2} A_{f} A_{B} \rho_{0}/(N\cb)$. We make use of the 
expression (\ref{prodamplitude}) derived in Appendix \ref{detailcalcul}
for the product of amplitudes $A_fA_B$ and obtain :
\begin{equation}
\Re S^{1} - 1 = - \frac{\pi}{(1+\gamma)N}  
\Re \tan \left( \pi \Delta_{f} - \frac{i\alpha}{2} \right)
\end{equation}
After expressing  $\alpha$ in terms of $q_0$ using  (\ref{RelationQ0Theta})
 as :
\begin{equation}
\tanh\left({\alpha  \over 2 } \right) = -\cot( \pi \Delta_{f})
\tan \left(\frac{(1-2q_{0})\pi}{2(1+ \gamma)}  \right)
\end{equation}
$\Re S^{1}$ coincides with the real part of (\ref{largeNS}). 
 
We now consider $\Im S^{1}$, for which we need to go beyond the 
scaling limit and use global properties of the Green's functions. 
First expressing $\Sigma_{f}$ as a convolution on the imaginary axis and
using 
$ \partial_{\omega }G_{0}(i\omega ) \rightarrow  - 2 i \pi \rho_{0}
 \delta (\omega )$ in the limit of a flat particle-hole symmetric band
 we obtain :
\begin{equation}\label{valG}
-i \rho_0 \gamma 2 \pi {\cal G} ( i \nu) = {\cal A} + {\cal B}(\nu)
\end{equation}
with the definitions :
\begin{eqnarray}
{\cal A}&=&
\int d \omega G_f (i  \omega ) \partial_{\omega} \Sigma_f(i \omega)\\
{\cal B}(\nu)&=& \int d \omega 
\left[  G_f (i  \omega + i \nu )\partial_{\omega }\Sigma_f(i \omega) -
G_f (i\omega)\partial_{\omega }\Sigma_f(i \omega)  \right]
\end{eqnarray}

In the limit of vanishing $\nu$,${\cal B}(\nu)$
can be calculated from the  scaling limit of $G_{f}$.
We obtain :
\begin{equation}\label{valB}
{\cal B}(\nu\rightarrow 0^+)
=\frac{\gamma}{1+\gamma}
\left( (-e^{\frac{-i\pi(1-2q_0)}{1+\gamma}}
+e^{-2i\pi\Delta_f})\frac {\pi}{\sin 2\pi \Delta_f}+i\pi \right)
\end{equation}
On the other hand, ${\cal  A}$ contains high frequency information 
that is lost in the scaling limit.
We find : 
\begin{equation}\label{valA}
{\cal A}=-{i\pi \gamma  \over 1+\gamma }(1-2q_{0})
\end{equation}
The details of these calculations are 
provided in Appendix \ref{appendixtmatrix}. 

Combining (\ref{valA},\ref{valB},\ref{valG},\ref{S1vsG}) we find agreement
with  the large $N$ limit of $S_{1}$ in Eq. (\ref{largeNS}).

For a dilute array of impurities (of concentration $n_{imp}$), the 
conduction electron-self energy is given by 
$\Sigma(\omega+i0^+)\simeq n_{imp} T(\omega)$, to lowest order 
in $n_{imp}$. As shown above, $T(\omega)$ is given in the 
large-N approach by the Fourier transform of $G_f(\tau)G_B(-\tau)$. 
The expansion (\ref{defsndfntechelle}) yields the long-time behaviour
$G_f(\tau)\sim A_f/\tau^{2\Delta_f} + A_f^{(2)}/\tau^{4\Delta_f}+\cdots$ and
$G_B(\tau)\sim A_B/\tau^{2\Delta_B} + A_B^{(2)}/\tau+\cdots$.
 From the fact that $2\Delta_f+2\Delta_B=1$, this implies:
$G_f(\tau)G_B(\tau)\sim 1/\tau + 1/\tau^{1+2\Delta_f}+\cdots$. Hence
the resistivity behaves as: 
\begin{equation}
\rho(T)  
\sim n_{imp}\rho_u \left( 
{{1-\Re S^1}\over{2}} - c T^{2\Delta_f}+\cdots \right)
\end{equation}
Where $\rho_u$ is the impurity resistivity in the unitary limit.
For the
same reasons as above, the Anderson model result would lead to an exponent
$2\Delta_B$ in the regime $\gamma<1$ \cite{CR}.

\subsection{The limit of a large number of channels 
($\gamma\rightarrow\infty$)}

We finally emphasize that all the expressions derived above greatly simplify 
in the limit of a large number of channels $\gamma\rightarrow\infty$. This is 
expected, since in this limit the non-Fermi liquid intermediate coupling 
fixed point becomes perturbatively accessible from the 
weak-coupling one \cite{NB,GAP}. 
The physics of the fixed point can be viewed as an almost free spin 
of ``size'' $Q=N q_{0}$ weakly coupled to the conduction electrons.
Indeed the large-$\gamma $ expansion
 of the entropy (\ref{Entropy}), the Green function $G_{f}$
 (\ref{fullGreenGenerales}) and the twist $\alpha $ (\ref{RelationQ0Theta})
 are : 

\begin{eqnarray}
\nonumber
S_{imp}&=& - \left(q_{0} \ln q_{0} + (1-q_{0}) \ln  (1-q_{0})  \right)
 - \frac{ \pi^{2}q_{0}(1-q_{0})}{6 \gamma^{2}} + 
\cdots \\
\nonumber
g_{f}(\tilde{\tau }) &=& 
{e^{\alpha (\tilde{\tau }-\frac{1}{2})}\over \cosh \frac{\alpha }{2} }
\left[1+\frac{1}{\gamma }\ln  \frac{\pi }{\sin \pi \tilde{\tau }}
+\cdots  \right]\\
\nonumber
\rho_{f}(\omega) &=&  
\frac{1}{T} \delta \left( \frac{\omega }{T}+\alpha  \right) + \cdots \\
 \alpha &=&  \ln \frac{q_{0}}{1-q_{0}} + \cdots 
\end{eqnarray}

and the leading terms in these expansions are given by the corresponding 
quantities for a free spin of ``size'' $Nq_{0}$.
Moreover the scattering matrix $S^{1}$ and resistivity have 
the following expansion :

\begin{eqnarray}
\nonumber
\Re S^{1}  &=& 1 - \frac{2\pi^{2} q_{0}(1-q_{0})}{N\gamma^{2}} 
\dots \,\, , \,\, 
\Im S^{1}  =  O\left(\frac{1}{\gamma^{3} }\right)\\
\rho (T=0) / (n_{imp}\rho_{u})  &=& \frac{1}{N} q_{0}(1-q_{0})
\frac{\pi^{2}}{\gamma^{2}} + \dots  
\end{eqnarray}

while the anomalous dimensions read 
$\Delta_{S}= 2\Delta_{f}= \frac{1}{\gamma } -\frac{1}{\gamma^{2}} + \dots $ 
, $2\Delta_{B}= 1 - \frac{1}{\gamma }  + \frac{1}{\gamma^{2}} + \dots $.

\section{Conclusion}

In this paper, we have focused on the {\it non-Fermi liquid overscreened
regime} of the
$\mbox{SU}(N)\times\mbox{SU}(K)$ multichannel Kondo model. 
This model has actually a wider range of possible behaviour, 
which become apparent when 
other kinds of representations of the impurity spin are considered. 
In a recent short paper \cite{OPAG}, two of us have studied 
{\it fully symmetric} representations corresponding to Young tableaus 
with a single line of $P$ boxes. 
(This amounts to consider {\it Schwinger bosons} in place of the 
Abrikosov fermions used in the present work). 
It was demonstrated that, in that case, a transition occurs as a 
function of the ``size'' $P$ of the impurity spin, from 
overscreening (for $P<K$) to underscreening (for $P>K$), with an 
exactly screened point in between ($P=K$). The large-N analysis of 
the overscreened regime $P<K$ is essentially identical to that 
presented in the present paper for antisymmetric representations. 

Obviously, an interesting open problem is to understand the physics 
of the model for more general impurity spin representations, 
involving both ``bosonic'' and ``fermionic'' degrees of freedom 
(corresponding respectively to the horizontal and vertical 
directions in the associated Young tableau). CFT methods are a 
precious guide in achieving this goal. In particular, the formulas 
and rules given in Appendices \ref{appendixstrongcoupling} and 
\ref{modsmatrix} allow for an easy derivation of the impurity 
$T=0$ residual entropy and zero-frequency T-matrix, using 
Affleck and Ludwig's fusion principle and the identification of 
these quantities in terms of modular S-matrices.

An open question which certainly deserves further studies is to 
identify which of these more general spin representations are such 
that a direct large-N solution of the model can be found. This 
question has obvious potential applications to the 
multi-impurity problem and Kondo lattice models. 

During the course of this study, we learned of a work by 
A.Jerez, N.Andrei and G.Zar{\'a}nd on the same model using the 
Bethe Ansatz method. 
Our results and conclusions agree when a comparison is possible (in 
particular for the impurity residual entropy and low-temperature behaviour 
of physical quantities).

\acknowledgements
We are most grateful to N.Andrei for numerous 
discussions on the connections between the various approaches. 
We also acknowledge discussions with S.Sachdev 
at an early stage of this 
work, about the importance of the thermal scaling functions. 
This work has been partly supported by a CNRS-NSF collaborative research 
grant NSF-INT-14273COOP. 
Work at Rutgers was also supported by grant NSF 9529138.

\appendix 
\section {The strong coupling state}
\label{appendixstrongcoupling}
We now describe in more details the proof of the statements in 
Sec.(\ref{strongcoupling}) about the nature and degeneracy of
 the strong-coupling state $R_{sc}$. For a general 
reference on the group theory material used in this appendix, 
the reader is referred {\it e.g} to Ref.\cite{Cornwell}. 
Let us note $\ny$ the number of electrons brought on the impurity site
and by $Y$ the Young tableau with $\ny$ boxes associated with the 
representation in which the 
conduction electrons on the impurity site combine. Because of the Pauli 
principle,  
{\sl the length  of any of its lines must be smaller than K} (and hence 
$\ny$ must be smaller than $NK$). 
Indeed, we must antisymmetrise the wave function separately for each flavour.

The Kondo energy is given by :
\begin{equation} 
E=J_K \sum_{\alpha \beta} S_{\alpha \beta} S'_{\beta \alpha}
\end{equation}
with
\begin{equation}
S_{\alpha\beta} = f^\dagger_{\alpha}f_{\beta} - {Q\over N}\delta_{\alpha\beta} 
 \,\,\,\, S'_{\alpha \beta}= c^\dagger_{\alpha} c_{\beta} 
\end{equation}
in which $f$ denotes the pseudo-fermion and $c$ the conduction electrons at
 the impurity site.
We can introduce the linear combinations : 

\begin{equation}
 T_{\alpha\beta}= {S_{\alpha\beta}+S_{\beta \alpha}\over \sqrt 2} 
  \,\,\,\,\,\,\, \alpha>\beta\,\,\,\,\,\,\,
  T_{\alpha\beta}= {S_{\alpha\beta}-S_{\beta \alpha}\over i\sqrt 2}
   \,\,\,\alpha<\beta\,\,\,\,\,\,\,
 T_{\alpha\beta}= S_{\alpha\alpha}   \,\,\, \alpha=\beta
\end{equation}

such that:
\begin{equation}
\sum_{\alpha \beta} S_{\alpha \beta} S_{\beta \alpha}= 
\sum_{\alpha\beta} T_{\alpha\beta}^2
\end{equation}
This leads to the following expression of the Kondo energy:
\begin{equation}
\label{energystrcou}
{2E\over J_K}= C_2(R_{sc}) - C_2(Y) - C_2(R)
\end{equation}
in which $C_2(Z)$ denotes the quadratic Casimir operator of the 
representation $Z$. The representation  
$R_{sc}$ is the specific component of $Y\otimes R$ associated with the 
bound-state formed by the impurity spin and the conduction electrons 
at strong-coupling. 
We recall that $R$ is a column of length $Q$ in this paper.

We have to minimize $E$ over all possible choices of $Y$ and of $R_{sc}$.

\begin{figure}
\fig{15cm}{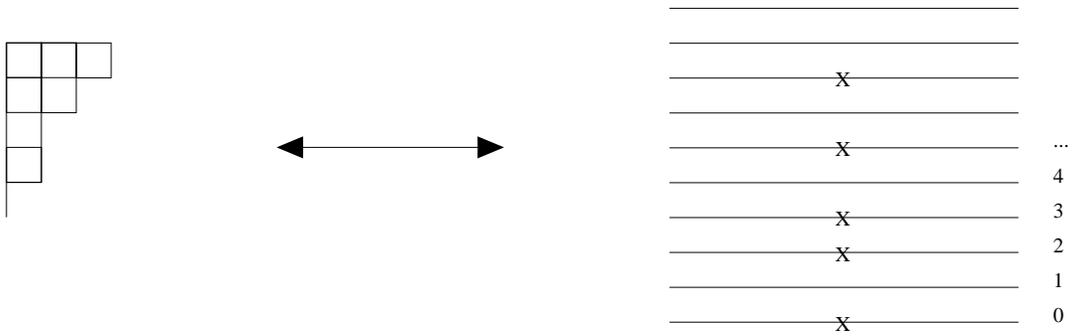}
\vskip 1cm
\caption{\label{DefEx}
An example of an $SU(N)$ Young tableau (for $N=5$) and its associated
 fermionic representation }
\end{figure}

First let us recall that for a general representation $Y$, $C_2$ is given by :
\begin{equation}
C_2= {1\over N} {}^t \vec n A^{-1} ({\vec n\over 2}+1)
\end{equation}
where $n_i \,\,(1\leq i \leq N-1)$ is the number of columns 
with length $i$ in the Young tableau Y and $A$ is the Cartan matrix 
of the SU(N) group \cite{Cornwell}.
Let us denote by $f_j \,\,(1\leq j \leq N)$ the length of the 
line $j$ in the tableau. Then we have
\begin{equation}
\label{formuleC}
C_2={1\over N} \left( {1\over 2} \sum _{j=1}^N (f_j-j+N)^2 - \left(
 {1\over 2N} \ny^2+ {N-1\over 2} \ny\right) - {N(N-1)(2N-1)\over 12}\right)
\end{equation}
with $\ny= \sum _{j=1}^N f_j$ is the number of boxes of $Y$. 
Note that with this definition, all $f_j$'s can be shifted by the same 
constant without changing the representation (this 
is because a column of length $N$ can
be removed  without changing the representation).
(\ref{formuleC}) can be given a simple interpretation in terms of $N$  
``particles'' occupying a set of fermionic levels. This interpretation was 
introduced in a slightly different form by Douglas  
\cite{Douglas}. 
Let $p_j=f_j-j+N$ be the position of the particle $j$. Because $Y$ is a 
Young tableau, the particles are ordered and cannot be on the same level.  
Fig.(\ref{DefEx}) gives an example of the construction of the 
diagram associated with a simple Young tableau. 
 
\begin{figure}
\fig{15cm}{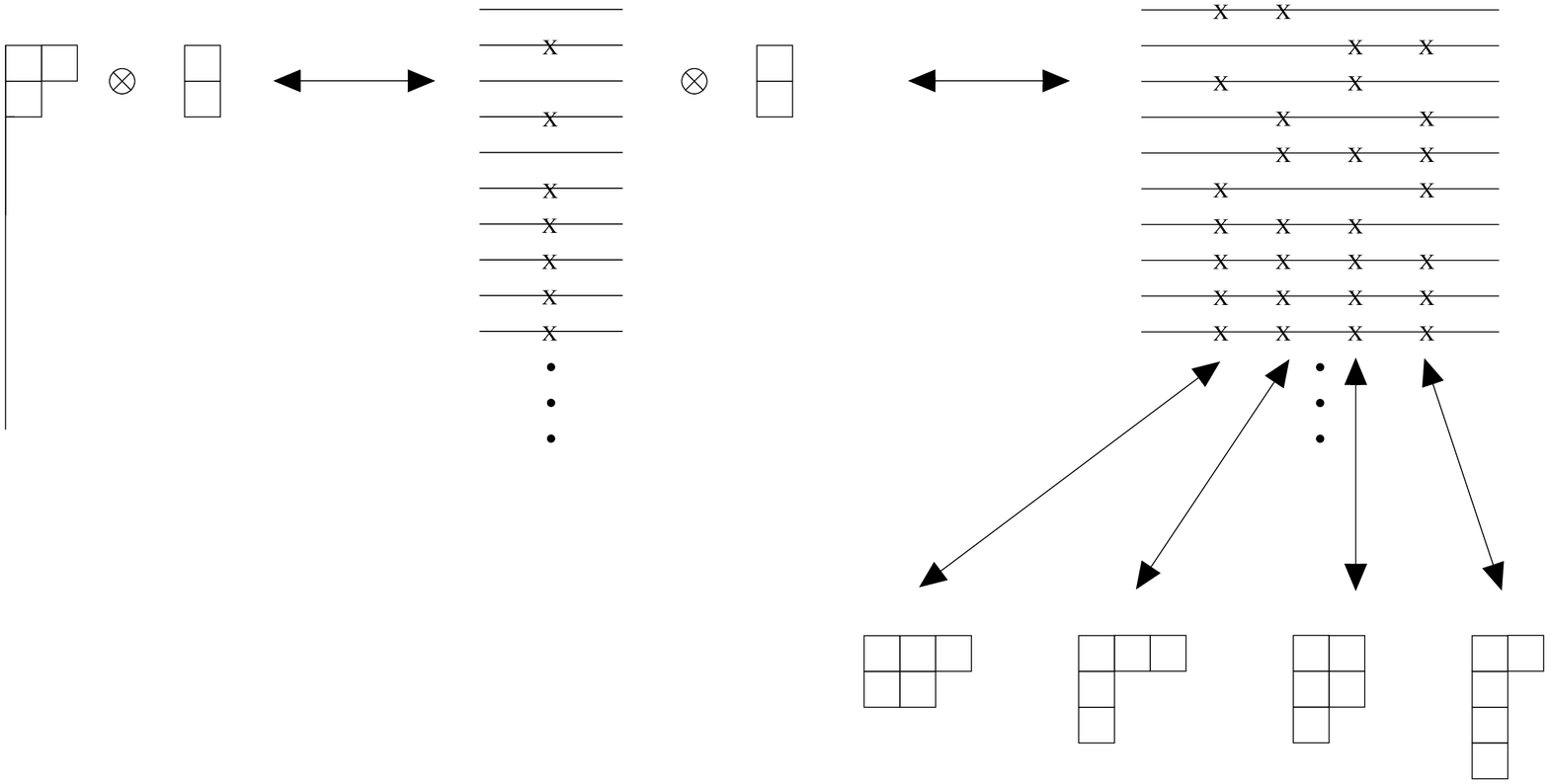} 
\vskip 1cm
\caption{ \label{CompoEx}
An example of the general composition rule explained in the text. 
We show the fermionic diagram associated with $Y$, 
the resulting fermionic diagrams and 
their transcription in terms of Young tableaus. 
($N$ is arbitrary in this example as evidenced by the \dots.)}
\end{figure}

A simple construction of all allowed Young tableaus appearing in the 
tensor product $Y\otimes R$ \cite{Cornwell} can be given in 
this fermionic language. Starting with the diagram associated with $Y$, 
we choose $Q$ particles and raise each of them by one level beginning with 
the one in the highest level. (We note that, in the fermionic interpretation, 
adding a box to line $i$ corresponds 
to raising the $i$-th particle up by one level). An example is given in  
Fig.(\ref{CompoEx}). 

Let us denote by $p_i$ the positions of the $N$ ``particles'' in $Y$, and 
by $p'_i$ the new positions in a given allowed component of $Y\otimes R$. 
The Kondo energy is given by:
\begin{equation}
\label{formuleC2}
E = {1\over 4N} \left( \sum_{j=1}^N ({p'}_i^2 - p_i^2) - 
{\ny\over 2 N} - {2N-1\over 8} - C_2(R)\right)
\end{equation}
The last two  terms are constant ($R$ is held fixed) and can be dropped in the 
minimisation process. 

The $p_i$'s can be decomposed in two sets: those for which $p'_i=p_i$ (we have 
$N-Q$ of them) and those for which $p'_i=p_i+1$ ($Q$ of them). 
Let us denote by $P$ the sum of the latter ones. We have:
\begin{equation}
E = {1\over 4N} \left(
Q+2P - {\ny\over 2 N} - {2N-1\over 8} - C_2(R)\right)
\end{equation}
Thus the lowest energy is achieved for the smallest possible value of $P$. 
Since a given shift $p\rightarrow p+1$ can only appear once in the sum  
(because double occupancies are forbidden and a given particle 
cannot be raised twice), 
the absolute minimum is obtained when we sum on all the lowest $Q$ shifts. 
This implies that the diagram associated with $Y$ has $Q$ particles on the 
$Q$ lowest levels (from 0 to $Q-1$) and none on the $Q$-th level.

The upper part of the diagram (above level $Q$) is then determined by 
the maximisation of $\ny$.   
Going back to the language of Young tableaus, the minimum is thus achieved 
when $Y$ is a rectangle of height $N-Q$ and width $K$, 
and $R_{sc}$ is given by the same tableau with the first column removed.

Two cases must thus be distinguished:
\begin{itemize}
\item For ($K=1$) and for arbitrary $N$ and $Q$, 
$R_{sc}$ is the trivial (singlet) representation (of dimension 
$d(R_{sc})=1$). 

\item For ($K\geq 2$, arbitrary $N$ and $Q$) the dimension $d(R_{sc})$ is 
is larger than the dimension of $R$. Indeed, 
denoting by $d_K(R_{sc})$ the dimension $R_{sc}$ for K-channels, we have 
the recursion relation (from the 
``hook law'' \cite{Georgi}): 
\begin{equation}
{d_{K+1}\over d_K}={(N+K)(N+K-1)\cdots (N+K-Q+1) \over (Q+K)(Q+K-1)\cdots 
(Q+K+1-Q)} >1
\end{equation}
(because $Q < N$). It increases with $K$. 
The $K=2$ case is just a column of length $N-Q$ which has the dimension
 of $R$. 
Moreover the inequality is strict for $K>2$.
\end{itemize}

\section{Construction of modular S-matrices}
\label{modsmatrix}
If two representations $R$ and $R'$ correspond to fermion configurations
with positions $\{p_1,\ldots,p_N\}$ and $\{p_1',\ldots,p_N'\}$ 
(See Appendix \ref{appendixstrongcoupling})
respectively, then the modular $S$-matrix element is :
\begin{equation}
S_{R,R'}=C_{N,K}e^{- \frac{2\pi iN\bar{p}\bar{p}'}{N+K}}
\det[e^{\frac{2\pi i p_i {p'}_j}{N+K}}]
\end{equation}
with $\bar{p}=\sum_i p_i/N$, $\bar{p'}=\sum_j p_j'/N$ and $C_{N,K}$ is a
 constant which depends only on $N$ and $K$.

Since for the trivial representation $0$, the $p$'s are the consequent 
integers
$0,1,\ldots,N-2,N-1$, $S_{0,R}$ involves a determinant of the form
\[ \left| \begin{array}{ccccc}
           1  & 1       & \cdots & 1           & 1       \\
           z_1& z_2     & \cdots &z_{N-1}      &z_N      \\
         z_1^2& z_2^2   & \cdots &z_{N-1}^2    &z_N^2    \\
        \vdots&\vdots   &        &\vdots       &\vdots   \\
     z_1^{N-2}&z_2^{N-2}& \cdots &z_{N-1}^{N-2}&z_N^{N-2}\\
     z_1^{N-1}&z_2^{N-1}& \cdots &z_{N-1}^{N-1}&z_N^{N-1}    
        \end{array} \right| \] 
where $z_j=e^{\frac{2\pi i p_j}{N+K}}$, $\{p_1,\ldots,p_N\}$ being the
positions of fermions corresponding to the representation $R$. 
This is just the Van Der Monde determinant $\Delta(z)=\prod_{i<j}(z_i-z_j)$

To calculate the $T$ matrix, we also need to know the $S$-matrix element
between  the fundamental representation $F$ and an arbitrary representation 
$R$. For $F$, the positions of the fermions are ${0,1,2,...,N-2,N}$.
Therefore $S_{F,R}$ involves
the determinant
\[ \left| \begin{array}{ccccc}
           1  & 1       & \cdots & 1           & 1       \\
           z_1& z_2     & \cdots &z_{N-1}      &z_N      \\
         z_1^2& z_2^2   & \cdots &z_{N-1}^2    &z_N^2    \\
        \vdots&\vdots   &        &\vdots       &\vdots   \\
     z_1^{N-2}&z_2^{N-2}& \cdots &z_{N-1}^{N-2}&z_N^{N-2}\\
         z_1^N&z_2^N    & \cdots &z_{N-1}^N    &z_N^N    
        \end{array} \right| \] 
This determinant has the same antisymmetry property in $z_i$'s as
 the Van Der Monde determinant. However, the present determinant is one
order higher than $\Delta(z)$ as a homogeneous polynomial in {z}'s.
A little reflection shows it is $(\sum_i z_i)\Delta(z)$.

Finally we find :
\begin{equation}
\frac{S_{F,R}}{S_{0,R}}= e^{- \frac{2\pi i\bar{p}}{N+K}} 
 \sum_{j=1}^{N} e^{\frac{2\pi i p_j}{N+K}}
\end{equation}

Using this formula we deduce (\ref{conformal}) from (\ref{conformal1}).

\section{Solution of the saddle point equations in the scaling regime}
\label{detailcalcul}
In this appendix we solve (\ref{sp}) in the scaling regime as explained in
section (\ref{spAnalysis1}) and obtain the scaled spectral densities and
 Green functions. 

\subsection{Scaling functions}

First we show that (\ref{ansatzCasGene}) is solution of the saddle point
 equations in the scaling regime. We deal with an arbitrary $q_0$.
Let us denote by  $\sigma_{f,B}$  the scaling function of the 
fermionic and bosonic self-energies : 
\begin{equation}
\Sigma_{f,B}(\tau) = A_{B,f}\, \beta^{-2\Delta_{B,f}-1} \sigma_{f,B}
({\tau\over \beta})
\end{equation}
$G_{0}$ is the local Green function for the conduction electron. Its 
density of states
does not depend on $T$. So its scaling form is :
\begin{equation}\label{scalingformG0}
G_{0}(\tau)= - {\rho_{0} \pi  \over \beta \sin {\pi \tau  \over\beta }}
\end{equation}
with $\rho_{0}=-{1 \over\pi  }\Im G_{0}(\omega=0)$.
Using this formula and (\ref{sp}), $\sigma_{f,B}$ are related to $g_{f,B}$.  
We insert the scaling form (\ref{scalg}) into (\ref{defsigma}).
 Matching the power in $\beta$ leads to $2\Delta_f + 2\Delta_B=1$ and :
\begin{eqnarray}
\label{eqscal}
\nonumber A_f^{-1}g_f^{-1}(i\bar \omega_n ) &=& \left(\lambda 
-\Sigma_f(i\omega_0 )\right)\beta^{2\Delta_B}
-A_B\left(\sigma_f(i\bar \omega_n)-
\sigma_f(i\bar \omega_0 )\right)\,\,\,,\,\,\, \\
A_B^{-1}g_B^{-1}(i\bar \nu_n) &=& \left({{1}\over{J}}-\Sigma_B(0)\right)
\beta^{2\Delta_f}-A_f\left(\sigma_B(i\bar \nu_n)-\sigma_B(0)\right)
\end{eqnarray}
with $i\bar \omega_n= i(2n+1)\pi$ and $i\bar \nu_n=i2n\pi$.
The term $i\omega_n$ in (\ref{defsigma}) vanishes in this scaling limit 
because  $\Delta_{f,B}<1$. 
We assume that at zero-temperature 
\begin{equation}
\label{cancel}
\lambda-\Sigma_f(0)={1\over J }-\Sigma_B(0)=0
\end{equation}
 so $\beta$ disappears of these equations at lower-order. 

We insert  our Ansatz into (\ref{eqscal}) with  the following Fourier
 transform formulas (which follow from \cite{Grad} 3.631) :

\begin{mathletters}
\label{fourier}
\begin{equation}
g_f(i \bar \omega_n) ={ (2\pi)^{\f} T^{\f-1} i (-1)^{n+1} \Gamma(1-\f) \over
\ca 
\Gamma\left(1- \Delta_{f} - {\bar \omega_n\over 2 \pi} + 
{i\alpha\over 2\pi}\right)
\Gamma\left(1- \Delta_{f} + {\bar \omega_n\over 2 \pi} -
 {i\alpha\over 2\pi}\right)
} 
\end{equation}
\begin{equation}
g_B(i\bar \nu_n) = { (2\pi)^{\B} T^{\B-1} (-1)^{n+1} 
\Gamma(1-\B)\over
\ca
\Gamma \left( 1- \Delta_{B} - {\bar \nu_n\over 2 \pi}
 + {i\alpha\over 2\pi}\right) 
\Gamma \left( 1- \Delta_{B} + {\bar \nu_n\over 2 \pi}
 - {i\alpha\over 2\pi}\right)
}  
\end{equation}
\begin{equation}
\sigma_f(i \bar \omega_n) -\sigma_f(i \bar \omega_0) =
{i \gamma \rho_0 (2\pi)^{\B+1} T^{\B} (-1)^{n+1} \Gamma(-\B)\over
\ca
\Gamma\left({1\over2} - \Delta_{B} - {\bar \omega_n\over 2 \pi}
 + {i\alpha\over 2\pi}\right)
\Gamma\left({1\over 2}- \Delta_{B} + {\bar \omega_n\over 2 \pi} 
 - {i\alpha\over 2\pi}\right)
 } - (n=0) 
\end{equation}
\begin{equation}
\sigma_B(i \bar \nu_n) -\sigma_B(0) =
{\rho_0 (2\pi)^{\f+1} T^{\f} (-1)^{n+1} \Gamma(-\f)\over 
\ca
\Gamma \left({1\over 2} - \Delta_{f} 
- {\bar \nu_n\over 2 \pi} + {i\alpha\over 2\pi}\right)
\Gamma \left({1\over 2} - \Delta_{f} 
+ {\bar \nu_n\over 2 \pi} - {i\alpha\over 2\pi}\right)
 } - (n=0) 
\end{equation}
\end{mathletters}

We see that (\ref{ansatzCasGene}) is solution of (\ref{eqscal}) provided that :
\begin{itemize}
\item The precise form of the cancelation (\ref{cancel}) at finite 
temperature is (at leading order in $T$) :
\begin{eqnarray}
{1\over J_K} - \Sigma_B(0) &=& 
{\rho_0 A_f (2\pi)^{\f+1} T^{\f}\Gamma(-\f)\over 
\ca
\Gamma \left( {1\over2} - \Delta_{f}  + {i\alpha\over 2\pi}\right)
\Gamma \left( {1\over 2}- \Delta_{f}  - {i\alpha\over 2\pi}\right) 
} \\
 \lambda - \Sigma_f(i\omega_0) &=& 
{i \gamma \rho_0 A_B (2\pi)^{\B+1} T^{\B}\Gamma(-\B)\over 
\ca
\Gamma \left( - \Delta_{B}  + {i\alpha\over 2\pi}\right) 
\Gamma \left( 1- \Delta_{B} - {i\alpha\over 2\pi}\right) 
}
\end{eqnarray}
\item (\ref{exponents}) is obeyed :
$ 2\Delta_f = {{1}\over{1+\gamma}}\,\,\,,\,\,\,
2\Delta_B = {{\gamma}\over{1+\gamma}}$
\item We have the relation between amplitudes :
\begin{equation}
\label{prodamplitude}
\B =-2\gamma A_f A_B \rho_0 \Gamma(1-\B)\Gamma(\B) 
{{\left|\sin
\left(\pi \Delta_{B} - {i\alpha\over 2}\right)\right|^2}
\over{\cb}}
\end{equation}
\end{itemize}

In our scaling forms, $\alpha $ is the same for the fermionic and the bosonic
function. One can check easily that for a  more general Ansatz
 with $\alpha_{f}$ and  $\alpha_{B}$ the saddle-point equations 
imply $\alpha_{f}= \alpha_{B}$.

\subsection{Spectral densities}

We calculate now the scaled spectral density from the above scaling function.
Denoting $\zeta=-1$ for fermions and $\zeta=1$ for bosons we have the general
formula.
\begin{equation}
G_{f,B}(\tau) = - \int_{-\infty}^{+\infty} {e^{-\tau \varepsilon}\over 
1 -\zeta e^{-\beta\varepsilon}} \rho_{f,B}(\varepsilon) d \varepsilon 
  \,\,\,\,\,  0\leq\tau\leq\beta
\end{equation}
In the scaling regime, we have to solve :
\begin{equation}
{{e^{\alpha (x-{1\over 2})}}\over{\ca}}
\left( {\pi \over \sin (\pi x)} \right)^{2\Delta_{f,B}} =
  \int_{-\infty}^{+\infty} {e^{-x u}\over 1 -\zeta e^{-u}} \phi_{f,B}(u)
 d u  \,\,\,\,\, 0\leq x \leq 1
\end{equation}

Setting $t=i(x-{1\over 2})$ we see it is sufficient to solve :
\begin{equation}
{{e^{- i\alpha t}}\over{\ca}}
\left( {\pi \over \cosh (\pi t)}\right)^{2\Delta_{f,B}} = 
 \int_{-\infty}^{+\infty} {e^{i t u}\over e^{u\over 2} -\zeta e^{-{u\over2}}}
 \phi_{f,B}(u) d u  \,\,\,\,\,  |\Im t|<{1\over 2}
\end{equation}
Due to the properties of Fourier transformation, we can just solve for 
$\alpha=0$,
and obtain the solution for arbitrary $\alpha$ with 
(\ref{DensiteSpectraleGenerales}).
With 
\begin{equation}
\label{fourier2}
\int_{-\infty}^{+\infty} d t
 \left( {\pi \over \cosh (\pi t)}\right)^\Delta e^{-itu}
 = (2\pi)^{\Delta-1} 
{
\Gamma\left({\Delta\over 2} + {iu\over 2 \pi}\right)
\Gamma\left({\Delta\over 2} - {iu\over 2 \pi}\right)
\over 
\Gamma(\Delta)}
\,\,\,\,\,\,\, 
 \left\{{ 0<\Delta<1 \atop u \hbox{ real}}\right.
\end{equation}
(see formula 3.313.2 of \cite{Grad}), we find the result given in the text
 (\ref{DensScalSym}) :
\begin{eqnarray}
\nonumber &\phi_f(\tilde{\omega},q_0=1/2)=
{{1}\over{\pi}}(2\pi)^{2\Delta_f-1}\cosh{{\tilde{\omega}}\over{2}}\,
{{\Gamma(\Delta_f+i{{\tilde{\omega}}\over{2\pi}})
\Gamma(\Delta_f-i{{\tilde{\omega}}\over{2\pi}})}
\over{\Gamma(2\Delta_f)}}\\
&\phi_B(\tilde{\omega},q_0=1/2)=
{{1}\over{\pi}}(2\pi)^{2\Delta_B-1}\sinh{{\tilde{\omega}}\over{2}}\,
{{\Gamma(\Delta_B+i{{\tilde{\omega}}\over{2\pi}})
\Gamma(\Delta_B-i{{\tilde{\omega}}\over{2\pi}})}
\over{\Gamma(2\Delta_B)}}
\end{eqnarray}

The asymptotic behaviour follows from formula 8.328 of \cite{Grad}.

We then derive the full Green function by taking a Hilbert transform :

\begin{equation}
g(z)=\int_{-\infty}^{+\infty} d x {\phi(x)\over z-x}
\end{equation}
We find (\ref{fullGreenGenerales}) using :   
\begin{itemize} 
\item      
The representation 
\begin{equation}
{1\over z-u} = -i \int_0^{+\infty} e^{i\lambda (z-u)} d \lambda 
\,\,\,\, \Im z>0
\end{equation}
\item The Fourier formula which inverses (\ref{fourier2})
\item The formula 
\begin{equation}
\label{fourier3}
\int_{0}^{+\infty} d x
{e^{iz x}\over 
 \left( \sinh {\pi x\over \beta}\right)^\Delta }
 = 2^{\Delta-1}{\beta\over \pi} 
 {
\Gamma\left({\Delta\over 2} - {i\beta z\over 2 \pi}\right)
\Gamma\left( 1- \Delta \right)
\over
\Gamma\left(1-{\Delta\over 2}-i{\beta z\over 2\pi}\right)
}
\,\,\,\,\,\,\,\,\,\,
\left\{{ 0<\Delta<1 \atop z \hbox{ real}}\right.
\end{equation}
which results from formula 3.112.1 of \cite{Grad}.
\end{itemize}

Finally, we comment on the treatment of the constraint equation 
(\ref{eqlambda}) in our derivation of the scaling funtions. 
The relation between $\alpha$ and $q_0$ has been derived from 
a Luttinger sum-rule, which holds at zero-temperature. So one 
may worry whether the scaling form does  satisfy the leading 
low-temperature corrections to the $T=0$ constraint equation. We show now 
that this is actually  the case. Starting from 
Eq.(\ref{eqlambda}) written as:
\begin{equation}\label{VerifContraint1}
\int_{\infty }^{\infty} d\omega \,\, \rho_{f}(\omega,T)n_{F}(\omega ) = q_{0}
\end{equation}
we substract the relation at $T=0$ and take into account the asymptotic
behaviour of $\phi_{f}$ given by Eq. (\ref{AsympPhif}) to obtain : 
\begin{equation}\label{VerifContraint2}
\int_{-\infty}^{0}dx \,\,\left(\phi_{f}(x) - \frac{e^{\frac{\alpha }{2}}
|x|^{2\Delta_{f}-1}}{\ca\Gamma (2\Delta_{f})} \right)
+ \int_{-\infty }^{\infty } d x \,\, \frac{\mbox{sgn} x \phi_{f}(x)}
{e^{|x|}+1}=0
\end{equation}
It is a rather strong constraint on the scaling function $\phi_f$ that 
this equation should hold, and it is satisfying that the explicit form 
obtained for $\phi_f$ does satisfy (\ref{VerifContraint2}). This proves 
that $\phi_f$ is really a solution
of the full system (\ref{sp},\ref{defsigma},\ref{eqlambda}) in the
scaling regime {\it at fixed} $q_{0}$.

\subsection{Higher-order terms in the scaling expansion}

Here, we give some indications on the derivation the expansion in  
(\ref{defsndfntechelle}). Let us start from the long-time expansion 
for $T_{K}^{-1}\ll \tau \ll \beta$:
\begin{equation}
G_{f}(\tau)\sim 
{A_{f} \over \tau^{2\Delta_f}}  + 
{A_f^{(2)}\over\tau^{\alpha} } \,\,\,\,\,\,
G_{B}(\tau)\sim {A_{B} \over \tau^{2\Delta_B}} + 
{A_B^{(2)}\over\tau^{\lambda} } \,\,\,\,\,\,
\end{equation}
in which $\alpha$ and $\lambda$ are exponents to be determined below. 
Then we have :
\begin{equation}
G_{f}(\omega) \sim A_{f} C_{\f-1} \omega^{2\Delta_f-1}  + 
A_f^{(2)}C_{\alpha -1} \omega^{
\alpha -1}
\end{equation}
with $C_\Delta= \int d t {e^{it}\over t^{\Delta+1}}$, and a 
similar expression  for $G_B$.
We can then deduce the expansions of $\Sigma_f$ and  
$\Sigma_B$, and insert them
into the saddle point equation. We find :
\begin{eqnarray}
\nonumber {\omega^{1-\f} \over A_f C_{\f-1}} - 
{A_f^{(2)} C_{\alpha -1} \over A_f^
2 C_{\f-1}^2} \,\omega^{\alpha +1-4\Delta_f} 
&=& \omega-{\gamma C_{\B} A_B \rho_0\over \pi
} \,\omega^{\B} - {\gamma C_{\lambda} A_B^{(2)} \rho_0\over \pi}
\,\omega^{\lambda}\\
{\omega^{1-\B} \over A_B C_{\B-1}} - {A_B^{(2)} C_{\lambda-1} 
\over A_B^2 C_{\B-1}^2}\,
\omega^{\lambda + 1-4\Delta_B} &=& - {C_{\f} A_f \rho_0\over \pi}\,
 \omega^{\f} - 
{C_{ \alpha} A_f^{(2)} \rho_0\over \pi}\, \omega^{\alpha} 
\end{eqnarray}
The first order yields:
\begin{equation}
\label{premierOrdre}
-\pi= \rho_0 A_f A_B C_{\f} C_{\B-1} =\gamma \rho_0 A_f A_B C_{\B} C_{\f-1}
\end{equation}
which, using:
\begin{equation} 
\label{relationC}
C_{\Delta-1} \propto  \Delta C_{\Delta}
\end{equation}
gives (\ref{exponents}) again. 
The second equation leads to $\lambda=\alpha + 1-4\Delta_f$.

First suppose $\lambda<1$ : 
in this case we must drop the $\omega$ term but we have
\begin{equation}
{ C_{\alpha -1} C_{\lambda-1}\over C_{\B-1}C_{\f-1}} =  
{ C_{\alpha} C_{\lambda}\over C_{\B}C_{\f}}
\end{equation}
which implies $\alpha=\f $ or $\alpha=\f-1$ (taking \ref{relationC} 
into account). 
So this possibility must  be rejected.
Finally we are lead to $\lambda=1$ and $\alpha=4\Delta_f$.

The higher order corrections can be dealt with in a similar manner. Restoring 
the scaling functions, this leads to (\ref{defsndfntechelle}). 

\section{Calculation of the residual entropy}\label{calculentropy}
\subsection{The formula of the free energy}
We first give a few more details on the regularisation in (\ref{free1}).
We will check that (\ref{regulbosons})
  is the right formula for the pseudo-boson.

In the following we will denote by $\mbox{Tr}_{\pm}$ the regularisation 
with $e^{i\omega_{n}0^{\pm}}$ and by $\mbox{Tr}_{sym}$ the regularisation
defined in (\ref{regulbosons}).  
We note that $\mbox{Tr}_{sym}=(\mbox{Tr}_{+} + \mbox{Tr}_{-})/2$ as can 
be checked explicitly 
using a spectral representation of the function to be summed.

Let us introduce the following notation 
for any quantity $A$ (function of $\lambda $) :
$\Delta_{\lambda } A =
 A^{\lambda } - A^{-\lambda } $.
 As the free energy is particle-hole symmetric,  we have : 
\begin{equation}\label{FreeEnergyIsPHSym}
-\lambda =\Delta_{\lambda } (T\mbox{Tr} \ln  G_{f}) -
 \gamma\Delta_{\lambda } (T\mbox{Tr} \ln  G_{B}) 
\end{equation}

Let us consider 
\begin{equation}
\phi (i\omega_{n}) = 
\ln \left(
{
i\omega_{n} + \lambda -\Sigma_{f}^{\lambda}(i\omega_{n}) \over
i\omega_{n} - \lambda -\Sigma_{f}^{-\lambda}(i\omega_{n}) 
}
\right)
\end{equation}
such as 
\begin{equation}
\Delta_{\lambda} (T\mbox{Tr}_{+} \ln  G_{f}) = - \phi(\tau =0^{-}) 
\end{equation}
As $\phi $ is  particle-hole symmetric, we have  
$\phi(\tau =0^{+})= - \phi(\tau =0^{-})$.
As its  asymptotic behaviour is 
$\phi (i\omega_{n}) \sim {2\lambda  \over i \omega_{n} }$,
its discontinuity is 
$\phi(\tau =0^{+}) - \phi(\tau =0^{-}) = -2\lambda $.

We obtain : 
\begin{equation}\label{relationlambda}
\Delta_{\lambda} (T\mbox{Tr}_{+} \ln  G_{f})= - \lambda 
\end{equation}

This implies that the bosonic term does not contribute to 
(\ref{FreeEnergyIsPHSym}). But there is an analogous relation for the boson :
we first calculate the discontinuity of $\Sigma_{B}$ from the saddle-point
equations, use an analogous function $\phi $
and obtain $\Delta_{\lambda} (T\mbox{Tr}_{\pm} \ln  G_{B})= \mp (1-2q_{0}) 
J /2$. 
So we find 
\begin{equation}
\Delta_{\lambda} (T\mbox{Tr}_{sym} \ln  G_{B})=0
\end{equation}
So we have checked  that (\ref{regulbosons}) 
is the right regularisation  for the bosonic term.

\subsection{Derivation of (\ref{free})}

We consider first the fermionic term. 
Let $G_{0}(i\omega_{n} )=\frac{1}{i\omega_{n} }$ be the Green function of
  free electrons. 
We have :
\begin{eqnarray}
\nonumber
T \mbox{Tr}_{+} \ln  G_{f} &=&
 -T \ln  2  - \frac{1}{\pi } \int_{\cal  R} d\omega \,\,
\left(\Im \ln G_{f} - \Im \ln G_{0} \right)n_{F}(\omega )\\
&=&-T \ln 2 + \frac{1}{\pi } \int_{\cal  R} d\omega \,\, 
\left(\arctan \frac{G_{f}'(x)}{G_{f}''(x)} + \frac{\pi }{2} - 
\pi \theta(-x) \right) n_F(\omega)\\
&=&  {1\over \pi} \int_{-\infty}^{+\infty} d \omega
 \left( \arctan \frac{G_{f}'(x)}{G_{f}''(x)}  -
 {\pi\over 2}\right) n_F(\omega) 
\end{eqnarray}

The bosonic term is obtained by an analogous calculation. 
In the particle hole symmetric case considered in the text
the three regularisations for the bosonic term are equivalent.
We have :
\begin{eqnarray}
\nonumber
- T \mbox{Tr}_{sym} \ln  G_{B} &=&
  - {1\over \pi} \int_{-\infty}^{+\infty} d \omega
\Im \ln (J G(\omega)) n_{B}(\omega) \\
&=&  - {1\over \pi} \int_{-\infty}^{+\infty} d \omega
 \arctan \frac{G''_{B}(x)}{G' _{B}(x)}  n_B(\omega)   
\end{eqnarray}

Finally we find the formula quoted in the text Eq. (\ref{free}).

\section{Some details of the T-matrix calculation}
\label{appendixtmatrix}
In this appendix, we calculate $\cal  A$ and ${\cal  B}(0^{+})$.
\subsection{Computation of ${\cal  A}$}

Using the definition of $\Sigma $ and introducing a oscillating  term
to regulate the 2 integrals, we have :
\begin{eqnarray}
{\cal  A}&=&
\int d \omega G_f (i  \omega ) \partial_{\omega} \Sigma_f(i \omega)\\
&=& i \int_{-\infty}^{\infty} d \omega G_{f}(i\omega)e^{i\omega 0^{+}} 
+  \int_{-\infty}^{\infty} d \omega \partial_{\omega } \ln G_{f}(i\omega)
 e^{i\omega 0^{+}} 
\end{eqnarray}

The first term is $2i\pi q_{0}$. Using (\ref{RelationQ0Theta}) and 
\begin{equation}\label{identite}
\int_{-\infty}^{\infty} d \omega { e^{i\omega 0^{+}} \over \omega  }=i\pi 
\end{equation}
(the integral is to be understood as a principal part), we have
with $\psi(z)=\ln(zG_{f}(z))$
 
\begin{eqnarray}
{\cal  A}&=& -2i\theta (1+\gamma) + \int_{i{\cal R}} dz
(\partial_{z}\psi)(z)\\
&=& -2i\theta (1+\gamma) - \lim_{\epsilon\rightarrow 0} 2i\Im  
( \psi (i\epsilon ) - \psi (i\infty))\\
&=& -2i\gamma \theta 
\end{eqnarray}

We used $\psi (i\infty)=0 $ and $\psi (i\epsilon )\sim \ln (A 
\epsilon^{2\Delta_{f}}) -i \theta  $ with $A$ a real constant.
Finally we find (\ref{valA}).

\subsection{Computation of ${\cal B}(0^{+}) $} 

\begin{equation}
{\cal  B}(\nu ) = i\int d\omega \left(G_{f}(i\omega +i\nu )-G_{f}(i\omega))
 \right)
- \int d\omega \left( G_{f}(i\omega +i\nu )-G_{f}(i\omega) \right)
\partial_{\omega }G_{f}^{-1}(i\omega )
\end{equation}

We  replace $G_{f}$ by the scaling function $g_{f}$.
The second term is of order 1 whereas the first is $O(T^{2\Delta_{f}})$
 and can be neglected.
We have then 
\begin{equation}
{\cal  B}(\nu ) = - \int dx \left(g_{f}(i(x+1)\tilde{\nu})- 
g_{f}(ix\tilde{\nu}) \right)\partial_{x} g_{f}^{-1}(i x \tilde{\nu })
\end{equation}
with $\tilde{\nu }=\frac{\nu }{T}$.
We want ${\cal  B}(\nu=0^{+},T=0)$ which is obtained by taking the limit 
$\tilde{\nu }\rightarrow +\infty$ in the previous scaling limit of ${\cal B}$.
To perform this limit we use 
$g_{f}(\bar z)=\overline{g_{f}(z)}$
and the following expansion for $g$ :
\begin{equation}\label{valueofA}
g_{f}(ix)\sim_{x\rightarrow +\infty } c A x^{2\Delta_{f}-1}
\hbox{\hskip 1cm with \hskip 1cm}
A=i \cosh \left(\frac{\alpha }{2}+i\pi \Delta_{f}  \right) 
\end{equation}
where $c$ is a real constant (Eq. (\ref{valueofA}) is obtained directly from
(\ref{fullGreenGenerales})).
We find :
\begin{eqnarray}
\nonumber-{\cal  B}(0^{+})&=&(1-2\Delta_f)\left[
-\int_{-\infty}^{-1}dx \frac{e^{i x 0^+}}{|x|^{2\Delta_f}|x+1|^{1-2\Delta_f}}
-\frac{A}{\bar A}\int_{-1}^{0}dx
 \frac{e^{i x 0^+}}{|x|^{2\Delta_f}(x+1)^{1-2\Delta_f}} + \right.\\
&&\left.\int_{0}^{\infty}dx
\frac{e^{i x 0^+}}
{x^{2\Delta_f}(x+1)^{1-2\Delta_f}}
-\int dx\frac{e^{i x 0^+}}{x} 
 \right]
\end{eqnarray}
The last term a principal part and is given by (\ref{identite}).
We then use the following identity :
\begin{eqnarray}
0&=&\int_{{\cal  R}+i0^{+}} dz \frac {e^{iz 0^+}}{z^{2\Delta_f}
(z+1)^{1-2\Delta_f}}\nonumber\\
 &=&
-\int_{-\infty}^{-1}dx
\frac{e^{i x 0^+}}
{|x|^{2\Delta_f}|x+1|^{1-2\Delta_f}}
+e^{-2i\pi\Delta_f}\int_{-1}^{0}dx
\frac{e^{i x 0^+}}
{|x|^{2\Delta_f}(x+1)^{1-2\Delta_f}}
\nonumber\\
& &
+\int_{0}^{\infty}dx
\frac{e^{i x 0^+}}
{x^{2\Delta_f}(x+1)^{1-2\Delta_f}}
\end{eqnarray}
We find 

\begin{eqnarray}
\nonumber {\cal  B}(0^{+})&=&(1-2\Delta_F)\left[
\left(\frac{A}{\bar A}+e^{-2i\pi\Delta_f} \right)
\int_{-1}^{0}d\omega \frac{e^{i x 0^+}}{|x|^{2\Delta_f}(x+1)^{1-2\Delta_f}}
+i\pi 
 \right]\\
&=& \frac{\gamma}{1+\gamma}\left[
\left(\frac{A}{\bar A}+e^{-2i\pi\Delta_f} \right)
\frac {\pi}{\sin 2\pi \Delta_f} +i\pi  \right]
\end{eqnarray}
A simple calculation with (\ref{valueofA}) shows :
\begin{equation}
{A \over\bar  A }= - e^{-\frac{i\pi }{1+\gamma }(1-2q_{0})}
\end{equation}
and we find (\ref{valB}).

\begin{figure}
$$\fig{6cm}{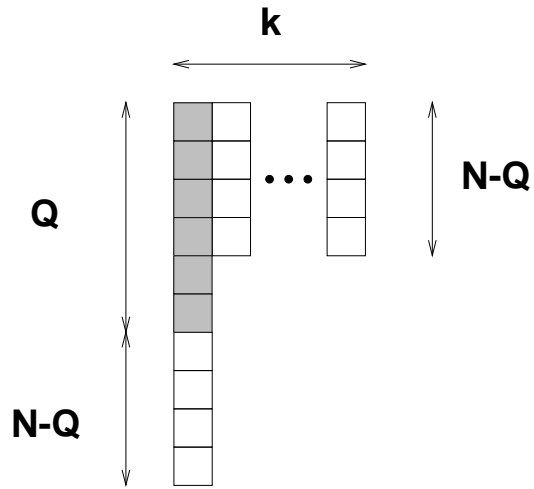}$$ 
\vskip 1cm
\caption{ \label{drawsc}
Young tableau corresponding to the strong-coupling state}
\end{figure}
\vskip 3cm

\vskip 3cm

\vskip 3cm

\vskip 3cm

\vskip 3cm


\begin{references}
\bibitem[*]{auth1}  Unit{\'e} propre du CNRS (UP 701) associ{\'e}e {\`a}
l'ENS  et {\`a} l'Universit{\'e} Paris-Sud.

\bibitem{NB} P. Nozi{\`e}res and A. Blandin,
{\sl J. Phys. (Paris)} {\bf 41}, 193 (1980).

\bibitem{CZ} For a recent review on non-Fermi liquid fixed points in
Kondo models, see D. L. Cox and A. Zawadowski, {\sl preprint
cond-mat/9704103}, to appear in Advances in Physics.

\bibitem{Moriond} See {\it e.g} {\it ``Correlated fermions and
transport in mesoscopic systems''}, T.Martin, G.Montambaux and
J. Tran Tanh Van eds, (Frontieres, 1996), and references therein.

\bibitem{AL} I.Affleck and A.W.W. Ludwig, {\sl Nucl. Phys. B}
{\bf 352}, 849 (1991); {\sl Nucl. Phys. B} {\bf 360}, 641 (1991).

\bibitem{AL2} I.Affleck and A.W.W. Ludwig, {\sl Phys. Rev. B} 
{\bf 48}, 7297 (1993).

\bibitem{ALent} I.Affleck and A.W.W. Ludwig, {\sl Phys. Rev. Lett.}
{\bf 67}, 161 (1991).

\bibitem{CR} D. L. Cox  and A. E. Ruckenstein,
{\sl Phys. Rev. Lett} {\bf 71}, 1613 (1993).

\bibitem{NCA} For a review, see N.E. Bickers, {\sl Rev. Mod. Phys.}
{\bf 59}, 845 (1987).

\bibitem{CS} B. Coqblin and J. R. Schrieffer, {\sl Phys. Rev.}
{\bf 185}, 847 (1969).

\bibitem{OPAG} O.Parcollet and A.Georges,
{\it ``Transition from overscreening to underscreening in the multichannel
Kondo model: exact solution at large-N''}
{\sl preprint cond-mat/9707337} to appear in Phys. Rev. Lett.

\bibitem{Smodulaire} V.Kac and D.Peterson, {\sl Adv. Math.} {\bf 53} 125
(1984); V.Kac and M. Wakimoto, {\sl Adv. Math.} {\bf 70} 156 (1988); see
also: E.J. Mlawer, S.G. Naculich, H.A. Riggs and H.J. Schnitzer,
{\sl Nucl. Phys.}, {\bf B352}, 863 (1991).

\bibitem{Douglas} M.R. Douglas {\sl preprint hep-th 9403119}.

\bibitem{EKetc} V. J. Emery and S. Kivelson,
{\sl Phys. Rev. B} {\bf 46}, 10812  (1992) ;
A.Sengupta and A.Georges, {\sl Phys.Rev.B} {\bf 49}, 10020 (1994) ;
D.G Clarke, T.Giamarchi and B.Shraiman
{\sl Phys.Rev.B} {\bf 48}, 7070 (1993).

\bibitem{foot1} In the exactly screened one-channel case, the $q_0$
dependence has been investigated both in large $N$ and by the Bethe Ansatz
method by P.Coleman and N.Andrei, {\sl J. Phys. C} {\bf 19} 3211, 1986. In
that case, Bose condensation of the auxiliary field $B(\tau)$ takes place
in the large-$N$ limit.

\bibitem{MH} E.M{\"u}ller-Hartmann, {\it Z. Phys. B} {\bf 57}, 281 (1984)

\bibitem{Tsvelik} See {\it e.g.} chapter 24 in A. M. Tsvelik
{\it ``Quantum Field Theory in Condensed Matter Physics''},
Cambridge University Press, 1995.

\bibitem{Subir} See also the recent work by S.Sachdev,
{\sl  cond-mat/9705206 and cond-mat/9705266}.

\bibitem{Cornwell} J.F. Cornwell { \it Group Theory in Physics}, vol.2,
  Academic Press  

\bibitem{AGD} Our proof is closest to that of Luttinger's theorem in
A.A. Abrikosov, L. P. Gorkov and I. E. Dzialoshinski,
{\it Methods of Quantum Field Theory in Statistical Physics},
revised edition by R. A. Silverman, Dover (New York), 1963.

\bibitem{Georgi}  H. Georgi {\it Lie Algebras in Particles Physics },
  Frontiers in Physics, Addison-Wesley, 
 
\bibitem{Grad}  I.S. Gradshteyn and I.M. Ryzhik {\it Table of integrals,
    series and products}, Academic Press, 1980

\bibitem{footentro} More precisely, we have found that inserting the 
scaling functions in $f_{imp}(T)-f_{imp}(0)$ and expanding to linear 
order in $T$ leads, for $q_0\neq 1/2$, to an {\it incorrect result}. 
Higher order terms apparently cannot be ignored in this expansion ! 
The situation is somewhat similar to $\mbox{Im}S^1$ in Sec.\ref{resistsec}.   
  
\bibitem{KC} A numerical
calculation of the impurity entropy within the {\it standard} NCA approach,
(which differs from ours) has recently appeared in 
 T.-S. Kim and D.L.Cox, Phys. Rev. B{\bf 55}, 12594 (1997).

\bibitem{GAP} J.Gan, N.Andrei and P.Coleman {\sl Phys. Rev. Lett} 
{\bf 70}, 686 (1993)

\end{references}
\end{document}